\documentclass[apj]{emulateapj}
\usepackage{apjfonts}
\usepackage{hhline}
\usepackage{amsmath}
\usepackage{graphicx}
\usepackage{natbib}
\usepackage{epstopdf} 
\usepackage{subfigure}
\usepackage{color}
\usepackage{comment}
\usepackage{longtable}

 \font\sevenrm=cmr7 scaled 1000

\newcommand{\OIII}{[O~{\sevenrm III}]}

\newcommand{\FeII}{Fe~{\sevenrm II}}

\newcommand{\SII}{[S~{\sevenrm II}]}
\newcommand{\OI}{[O~{\sevenrm I}]}

\newcommand{\NII}{[N~{\sevenrm II}]}
\newcommand{\Hb}{H$\beta$}
\newcommand{\Ha}{H$\alpha$}

\newcommand{\kms}{km s$^{-1}$}

\begin{document}
\title{Positive and negative feedback of AGN outflows in NGC 5728}
\author{Jaejin Shin$^{1}$}
\author{Jong-Hak Woo$^{1}$\altaffilmark{,4}}
\author{Aeree Chung$^{2}$}
\author{Junhyun Baek$^{2}$}
\author{Kyuhyoun Cho$^{1}$}
\author{Daeun Kang$^{1}$}
\author{Hyun-Jin Bae$^{1,3}$}

\affil{
$^1$Astronomy Program, Department of Physics and Astronomy, 
Seoul National University, Seoul, 08826, Republic of Korea\\
$^2$Department of Astronomy, Yonsei University, Seoul 03722, Republic of Korea\\
$^3$Department of Medicine, University of Ulsan College of Medicine, Seoul 05505, Republic Of Korea
}

\altaffiltext{4}{Author to whom any correspondence should be addressed}

\begin{abstract}
We present a spatially-resolved analysis of ionized and molecular gas in a nearby Seyfert 2 galaxy NGC 5728,
using the VLT/MUSE and ALMA data. We find ionized gas outflows out to $\sim$2 kpc scales,
which encounter the star formation ring at 1 kpc radius. 
The star formation rate of the encountering region is significantly high ($\sim$1.8 M$_{\odot}/\rm yr/kpc^2$) 
compared to other regions in the ring. In contrast, the CO (2-1) emission is significantly weaker by a factor of $\sim$3.5,
indicating very high star formation efficiency. These results support the positive feedback scenario that
the AGN-driven outflows compress the ISM in the ring, enhancing the star formation activity. 
In addition, we detect outflow regions outside of spiral arms, in which gas is likely to be removed from the spiral arms
and no clear sign of star formation is detected. 
The overall impact of AGN outflows on the global star formation in NGC 5728 is limited, suggesting the feedback of the
low-luminosity AGN is insignificant. 
\end{abstract}

\section{INTRODUCTION} \label{section:intro}

\begin{figure*}{}
\includegraphics[width = 0.94\textwidth]{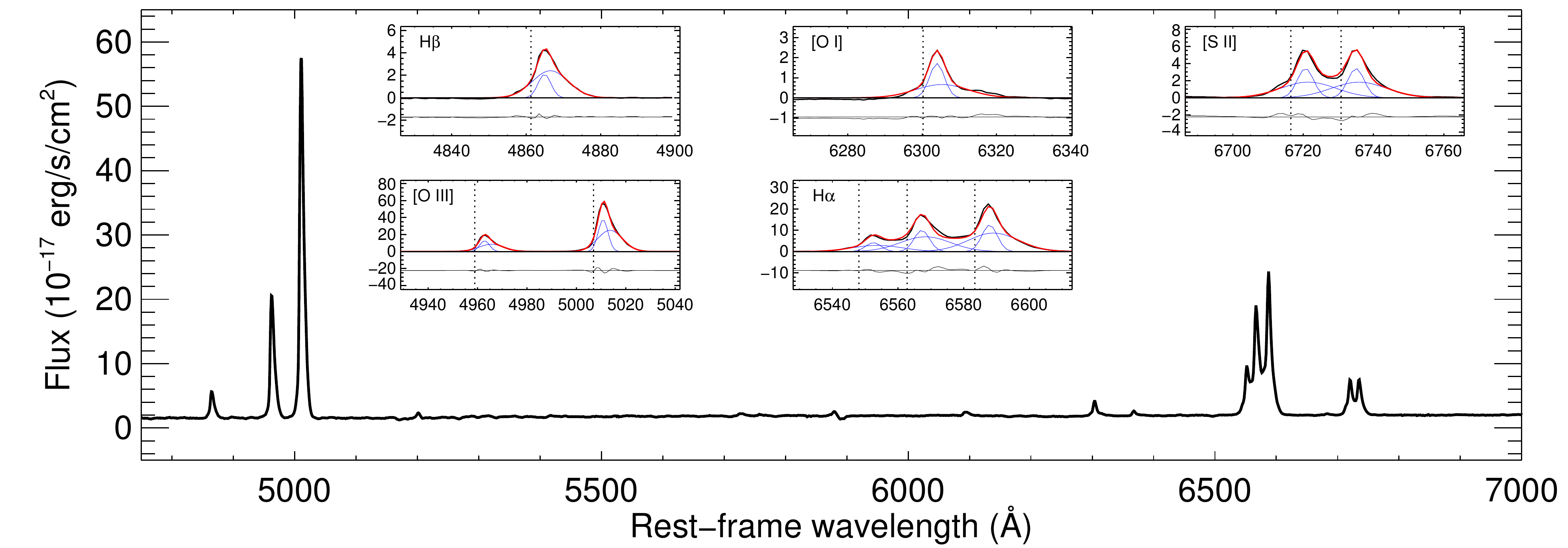}
\caption{
AGN spectrum extracted from the central 0.2 \arcsec $\times$ 0.2 \arcsec region. In the inset boxes, stellar subtracted spectrum (black), 
the combined model (red), and Gaussian components (blue) of each emission line from \Hb\ to \SII\ 
are presented, respectively. \\
\label{fig:allspec1}}
\end{figure*} 

Active galactic nuclei (AGNs) are one of the key components in understanding galaxy formation and evolution as they release large amount of energy via gas outflows or radio jets, affecting star formation in galaxies as well as the properties of intergalactic medium \citep[][and references therein]{Fabian2012,Harrison2017}.
In the AGN feedback scenarios, the energy from AGNs suppresses star formation by preventing gas from cooling (e.g., cooling flow problem, \citealt{Fabian1994}) or sweeping out gas from their host galaxies \citep[i.e., negative feedback, e.g.,][]{Silk1998,Fabian2012,Harrison2017}.
On the other hand, AGN activities can also compress gas and trigger star formation in other circumstances, e.g., high density regions \citep[i.e., positive feedback, e.g.,][]{Silk2013,Zubovas2013}. Thus, the nature of AGN feedback is complex since AGNs play a role in suppressing as well as enhancing star formation \citep{Zinn2013,Cresci2015a,Carniani2016,Zubovas2017}.
{ Star formation has been recently detected inside gas outflows, implying that gas outflows trigger star formation \citep{Maiolino2017,Gallagher2019}.} 

As a strong emission line in the optical wavelength range, the \OIII\ 5007\AA\ line has been popularly used as a tracer of ionized gas outflows. Various studies based on large survey data showed that gas outflows are prevalent in AGNs \citep[e.g.,][]{Mullaney2013,Bae2014,Woo2016,Rakshit2018}.
Nonetheless, the overall impact of AGN outflows on star formation is unclear, particularly when the spatially-resolved data are not available. Using $\sim$110,000 AGNs and star forming galaxies from the Sloan Digital Sky Survey (SDSS), for example, \cite{Woo2017} found that specific star formation rate of strong outflow AGNs is comparable to that of main-sequence star forming galaxies, suggesting no evidence of instantaneous negative feedback. 

The integral field spectroscopy has enabled the detailed investigation of the spatial connection between gas outflows and star formation. Using the spatially-resolved information of AGN gas outflow, a number of studies found that mass outflow rate is much higher (by $\sim$3-1400 times) than mass accretion rate \citep{Barbosa2009,Riffel2009,Storchi-Bergmann2010, Muller-Sanchez2011, Bae2017,Humire2018,Revalski2018,Durre2019} or star formation rate \citep[SFR;][]{ForsterSchreiber2014,Harrison2014}, indicating that gas removal is efficient. 
While the negative feedback is expected from the large amount of outflowing gas, the instantaneous feedback effect has not been well observed \citep[see e.g.,][]{Karouzos2016a,Karouzos2016b,Bae2017}.

Spatial anti-correlation between ionized gas outflows and star formation has beed detected in individual objects, suggesting  the negative feedback \citep{Cano-Diaz2012,Cresci2015a,Carniani2016}. On the other hand, \citet{Cresci2015a} and \citet{Carniani2016} found star formation activity at the edge of outflows, suggesting both positive and negative feedback for given objects. Interestingly, \cite{Cresci2015b} reported star forming regions where an AGN-driven gas outflow encounters a dust lane in a Seyfert 2 galaxy, NGC 5643, suggesting that gas outflows may trigger star formation in dense regions. 

Along with ionized gas, molecular gas provides crucial information on the outflows and their connection to star formation. 
Observational studies revealed the cold molecular gas outflows (i.e., CO molecules) due to AGN and the suppression of star formation \citep[e.g.,][]{Feruglio2010,Veilleux2013,Cicone2014,Fiore2017,Fluetsch2019}. Based on the Atacama Large Millimeter/submillimeter Array (ALMA) data, several studies reported that the spatially resolved kinematics of the molecular gas are consistent with that of the ionized gas, indicating both molecular and ionized gas are under influence of AGN \citep[e.g.,][]{Garcia-Burillo2014,Zschaechner2016,Slater2019}. { However, the cold molecular gas outflows have been investigated only for a small number of nearby AGNs based on spatially resolved observations.} 
On the other hand, warm molecular gas ($T\sim10^{3}$ K) traced by, for example, H$_{2}\lambda$2.1218$\rm \mu m$ does not usually show outflow signatures in nearby AGNs \citep[e.g.,][]{Riffel2013,Davies2014,Schonell2019}. It is possible that warm molecular gas has been destroyed due to the radiation from AGN \citep[e.g.,][]{Schonell2019}.

The various previous results based on tracers of ionized and molecular gas outflows showed that the effect of AGN-driven gas outflows is diverse and complex. In particular, it is important to investigate gas in multi-phase to fully understand the impact of the outflows. Detailed studies with a population of AGNs are required to unravel the nature of AGN feedback in galaxy evolution. 

In this paper, we focus on a nearby Seyfert 2 galaxy NGC 5728 at a distance of 40.3 Mpc, 
in order to investigate the connection between AGN outflows and star formation.
This galaxy presents a star formation ring at a $\sim$1 kpc radius, while ionized gas outflows in a biconical shape as well as one-sided radio jet are detected \citep{Schommer1988,Wilson1993,Son2009a,Davies2016,Durre2018,Durre2019}.
A particular interest is that the biconical gas outflows intersect with the star formation ring \citep[e.g.,][]{Wilson1993,Durre2018,Durre2019}, which is the densest region in the host galaxy \citep[similar to the dust lane of NGC 5643 of][]{Cresci2015b}, providing a good testbed for investigating AGN feedback via gas outflows. 
Recently, \cite{Durre2018,Durre2019} investigated the spatially-resolved kinematics of the ionized gas in the nuclear region of NGC 5728, reporting that a substantial amount of gas (38 M$_{\odot}$ yr$^{-1}$) is being removed from the nuclear region due to the powerful AGN gas outflows. They also reported no strong spatial relation between the radio jet and the supernova remnants in the star formation ring. However, the larger scale outflows and their impact on the star formation ring has not been perviously explored. Thus, we will focus on the outflow kinematics in the $\sim$6 kpc $\times$ $\sim$6 kpc scales and the connection between the gas outflows and star formation, particularly in the star forming ring, using the spatially resolved measurements based on the VLT/MUSE and ALMA data. 
In section 2, we describe the data and data preparation. The analysis is described in section 3, 
and we present our results and discussion in Section 4 and 5. We adopt a cosmology of 
$H_{\rm 0}= 70$ km  s$^{-1}$ Mpc$^{-1}$, $\Omega_{\Lambda}=0.7$ and $\Omega_{\rm m}=0.3$.  \\
 
\begin{figure*}{}
\includegraphics[width = 0.96\textwidth]{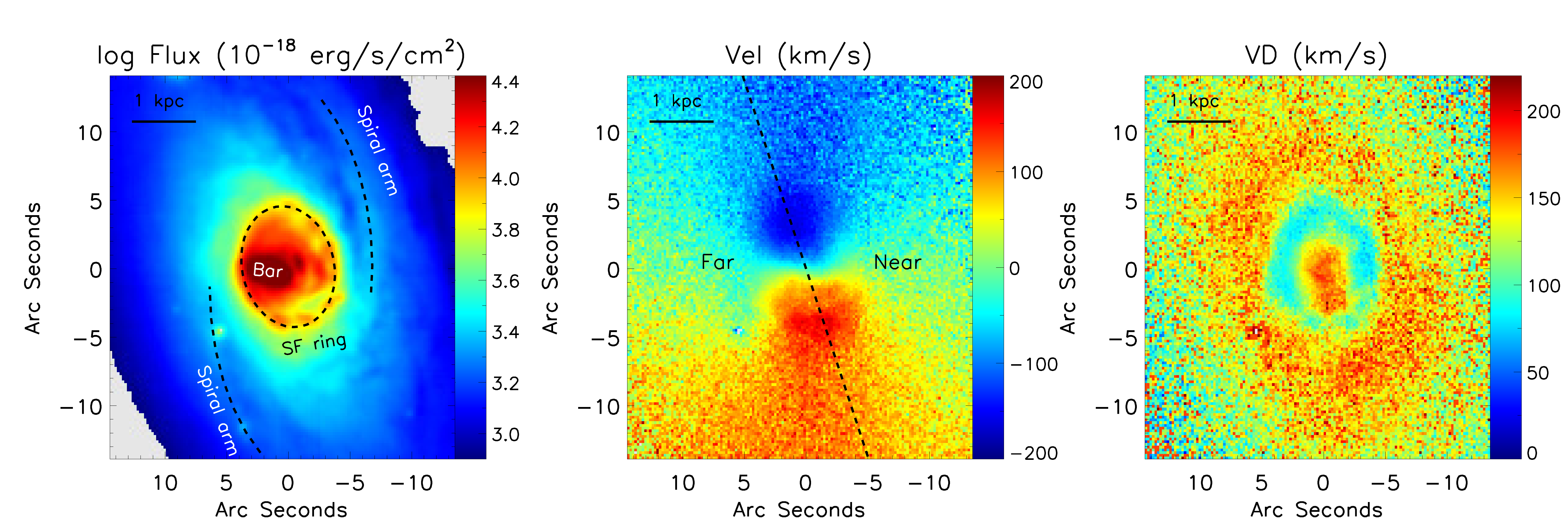}
\caption{
Flux, velocity, and velocity dispersion maps of the stellar component. { The flux was integrated from 4800 to 6800\AA.} The flux map shows a star formation ring, spiral arms, and a bar. In the velocity map, blue and red represent approaching (blueshift) and receding velocities (redshift), respectively. { Black dashed line indicates the
major axis of the star formation ring and divides far-side (SE) and near-side (NW).}\\
\label{fig:allspec1}}
\end{figure*} 

\begin{figure*}{}
\includegraphics[width = 0.92\textwidth]{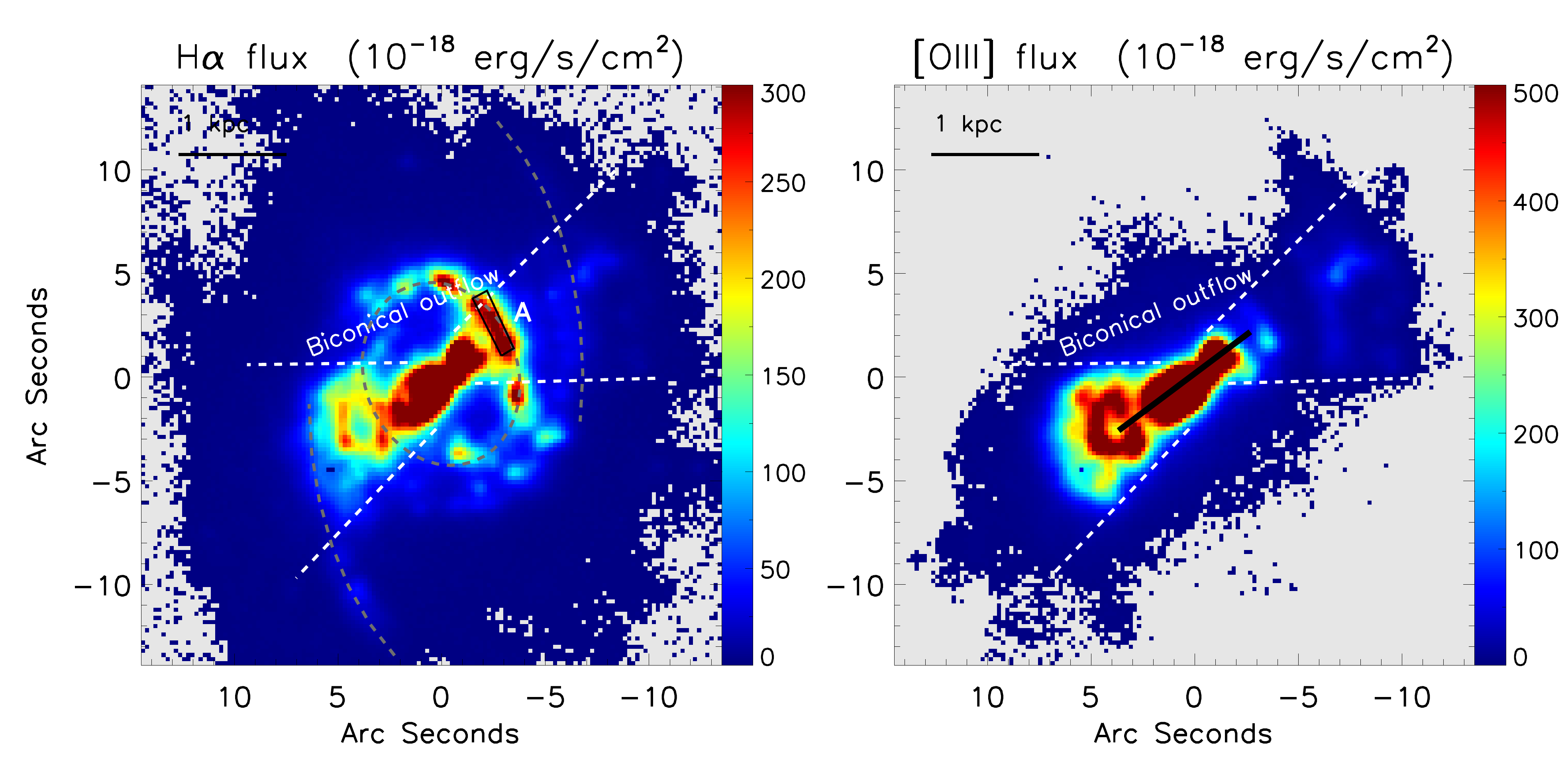}
\caption{
Flux maps of \Ha\ (left) and \OIII\ (right). Biconical gas outflows are represented with white dashed lines.
Gas outflows and the ring encounters in Region A (black box). Black solid line in the right panel indicates a pseudo-slit with the length of 8\arcsec\ 
to extract a one dimensional radial profile of flux, velocity, and velocity dispersion of \OIII, which is used in Section 4.4.
\label{fig:allspec1}}
\end{figure*} 

\section{Data}\label{section:Sample}
NGC 5728 was observed with the VLT/MUSE in 2016 Apr 3 and Jun 3 as a part of the Time 
Inference with MUSE in Extragalactic Rings (TIMER) survey 
\citep[Observation ID: 097.B-0640 (A), PI: Gadotti][]{Gadotti2019}. 
The observation was divided into 12 exposures, resulting in a total exposure time of 1.6 hrs.
The start seeing during the observing night ranged 0.56-0.91\arcsec.\ 
The data was retrieved from the ESO archive and reduced with the standard ESO reduction pipeline, 
ESOREFLEX (MUSE version of 2.4.2). The resulting seeing size for the combined cube is 0.66\arcsec.

To compare with the MUSE data, we also utilized the ALMA archival data of the 12 CO (2-1) line from Project 2015.1.00086 (P.I. N. Nagar). The raw data were re-calibrated and re-imaged using the pipeline, CASA 4.7.0. Phase, bandpass and amplitude were calibrated using J1448-1620, J1517-2422 and Titan, respectively. The channel was sampled into 10 km $\rm s^{-1}$ width and the robust weighting was set to be the robust 0 to optimize the data in sensitivity and resolution. The synthesized beam of the final cube is 0.55\arcsec $\times$ 0.47 \arcsec with a 1-$\sigma$ rms of $\sim$0.65 mJy per beam per channel. 
\\

\section{Analysis} \label{section:analysis}
\subsection{MUSE analysis} \label{section:anlysis}
For the MUSE datacube, we performed a spectral fitting analysis by focusing the central 30\arcsec$\times$30\arcsec, 
where stellar continuum or emission lines are clearly visible. In the analysis, we used the spectral window from $\sim$4800 to 6800\AA, which 
covers the main optical emission lines { (i.e., \Hb, \OIII$\lambda$4959,5007, \OI$\lambda$6300  \Ha, \NII$\lambda$$\lambda$6548,6583, and \SII$\lambda$$\lambda$6716,6731)}. 

First, we fitted and subtracted the stellar continuum using the Penalized Pixel-Fitting code \citep[][]{Cappellari2017} with 
47 ages (from 60 Myr to 12.6 Gyr) and solar metallicity of E-MILES templates \citep{Vazdekis2016}, { which is widely used as it provides large dynamic ranges of metallicity and age.}
In the stellar continuum fitting, we masked visible emission lines. 
Second, we fitted emission lines, which satisfy the amplitude-to-noise (A/N) ratio 
larger than 3, using the MPFIT package \citep{Markwardt2009}. 
To reproduce the observed emission lines, we adopted single- or double- Gaussian model as a line profile.
The double Gaussian model was considered only if the A/N ratio of each Gaussian component in the fitting result
is larger than 3, and the number of Gaussian components was determined depending on the chi-squares of the fitting results.
Most fitting parameters are given to be free except for the \OIII, \SII, and \NII. 
We tied the velocity and velocity dispersion for the \OIII, \SII, and \NII\ doublet, respectively. 
In the case of \NII6548 and \NII6583, we set flux ratio as 3. 
{ From the best-fit models, we measured flux, velocity, and velocity dispersion (VD) for each line (see Figure~1). Note that we did not separate the broad component from the narrow component in measuring outflow kinematics, since even the narrow component showed non-gravitational kinematics (see the narrow component of each line at the center in Figure~1).}
The uncertainties were measured based on Monte-Carlo simulations with 100 mock spectra, which were used in
comparison with the model predictions (see Section 4.4).

\subsection{ALMA analysis} \label{section:anlysis}
In order to investigate molecular gas distribution compared to that of ionized gas, we made a CO (2-1) line 
intensity map only with the data above $\sim$5$\sigma$ in order to be conservative. 
Using the line intensity, we estimated total molecular gas mass ($M_{\rm H2}$) 
by adopting Eq. 3 of \cite{Bolatto2013} and the same conversion factor as Milky Way, 
i.e. Xco = 2 $\times$ $10^{20} \rm cm^{-2}\ (K \ km^{-1})^{-1}$ \citep{Bolatto2013}. The mean CO (2-1) to CO (1-0) line ratio was assumed to be 
$\sim$0.8 \citep{Braine1992}. 
We note that it is arguable whether our assumptions for Xco and CO (2-1) to (1-0) line ratio are applicable for AGN hosts including this case. 
For example, Xco is measured to be 0.2-0.4 $\times$ $10^{20} \rm cm^{-2}\ (K \ km^{-1})^{-1}$ among the AGN population \citep{Bolatto2013}.
 CO (2-1) to (1-0) line ratio is also known to vary depending on the environment  \citep{Leroy2009}.
While investigating more reliable Xco and CO (2-1) to (1-0) line ratio is beyond the scope of this work, we would like to emphasize that all comparisons are made relatively within the area of interest, and hence our interpretation is robust regardless of the choice of those values.

\begin{figure*}{}
\includegraphics[width = 0.9\textwidth]{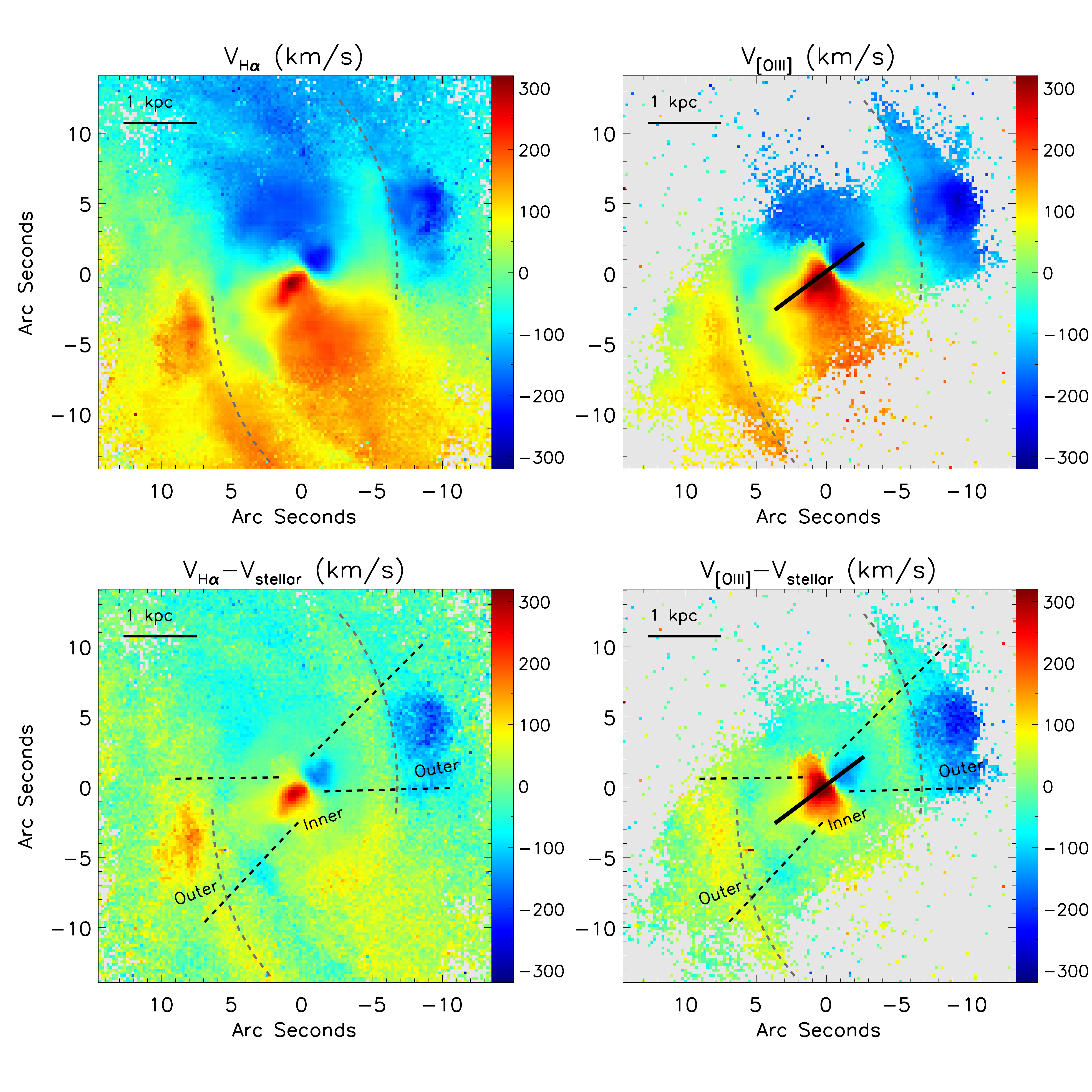}
\caption{
Upper panels: velocity maps of \Ha\ (left) and \OIII\ (right) with respect to the systemic velocity of NGC 5728.
Lower panels: maps of the relative velocity of \Ha\ (left) and \OIII\ (right), after subtracting stellar velocity in each spaxel 
(i.e., $V_{Ha}-V_{stellar}$ and $V_{[OIII]}-V_{stellar}$).
Spiral arms and biconical outflows are denoted with gray dashed lines and black dashed lines, respectively, as presented in Figure~3.
Black solid lines in the right panel are same as in Figure~3.
\label{fig:allspec1}}
\end{figure*} 

\begin{figure*}{}
\includegraphics[width = 0.9\textwidth]{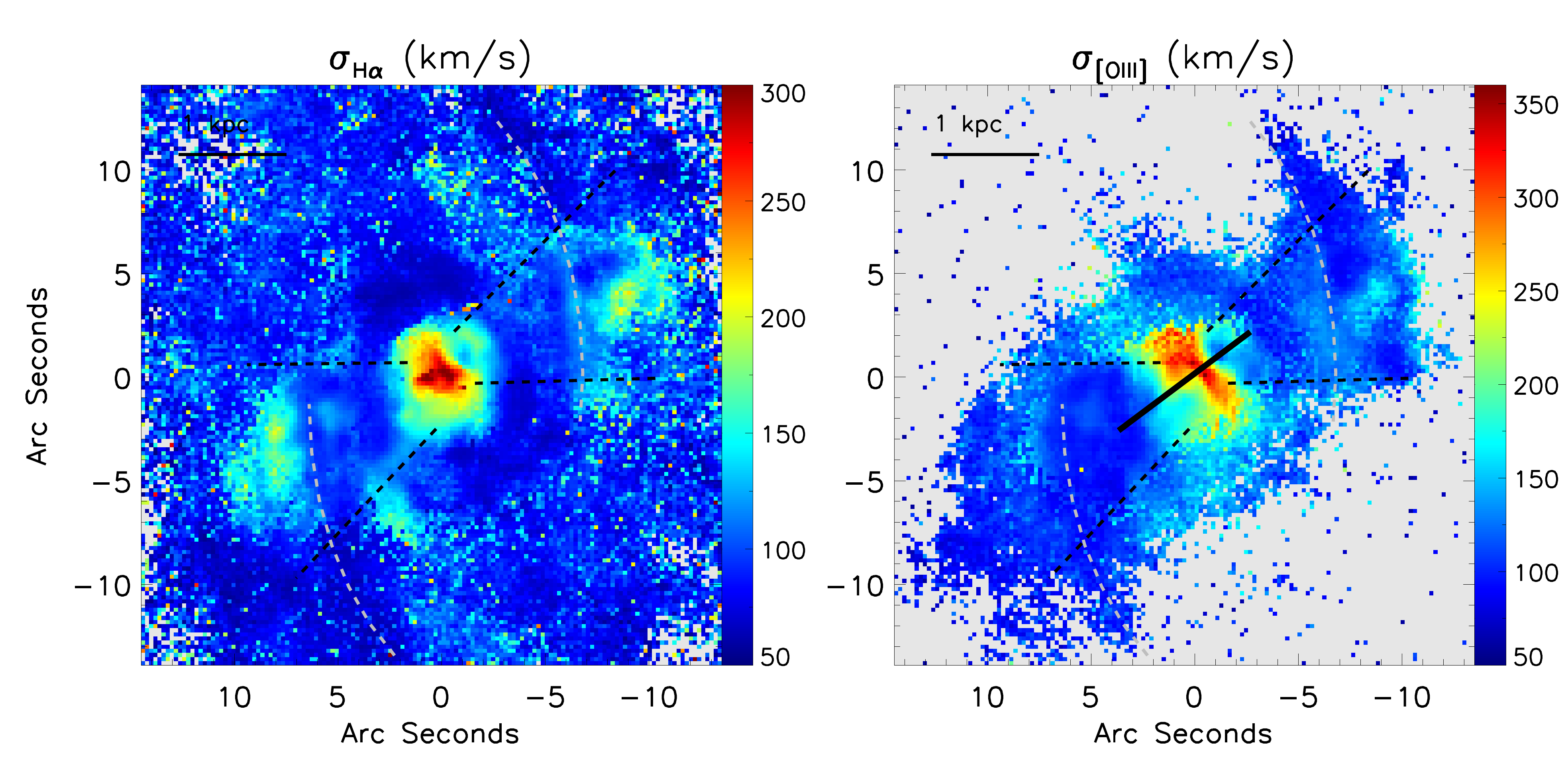}
\caption{
Velocity dispersion maps of \Ha\ (left) and \OIII\ (right). Spiral arms and biconical outflows are denoted with gray dashed lines and black dashed lines, respectively, as presented in Figure~3. Gray and black dashed lines and black solid line are same as in Figure~3.
\label{fig:allspec1}}
\end{figure*} 

\section{Result} \label{section:result}
\subsection{Stellar component}\label{astrometry correction}

In Figure~2, we present the spatial distributions of the flux, velocity, and velocity dispersion of the stellar component, 
which were measured based on the stellar continuum over the wavelength range of 4800-6800\AA.
The flux map shows interesting structures: 1) a star formation ring, 2) two spiral arms, and 3) a bar. 
The spiral arms may indicate the connection between the star formation ring and the large scale structure of the galaxy.
The bar was previously reported by \cite{Wilson1993} and also fully detected in the NIR observations by \cite{Ensellem2001}. 

The velocity map clearly shows a rotation pattern, of which the northern part is approaching while the southern part is receding with the maximum velocity, -160 \kms\ and 160 \kms, respectively. { Interestingly, the rotation pattern seems to be faster in the central region within 1 kpc compared to that in large scale (i.e., >1 kpc).
This result may suggest the presence of a compact disk at the central region of NGC 5728. }
{ Note that we determine the orientation of the disk as the SE region is far-side, while the NW region is near-side, based on the stellar kinematics assuming that the spiral arms are trailing. The same geometry was constrained by \cite{Son2009a} based on the information of the star formation ring although it is possible that the large scale disk and the star formation ring can be tilted from each other.} The velocity dispersion slightly decreases outwards with an average $\sim$150 \kms, except for the location of the star formation ring, where
stellar velocity dispersion is much lower with an average $\sim$100 \kms. Note that
the measured stellar kinematics is consistent with that of \cite{Durre2019}. \\

\subsection{Ionized gas}\label{astrometry correction}
In this section, we present the spatial distributions of flux, velocity, and velocity dispersion, which were measured from the \Ha\ and \OIII\ emission lines, respectively, to investigate AGN outflows. 

\subsubsection{Flux distribution}\label{astrometry correction}

The flux maps of \Ha\ and \OIII\ are presented in Figure~3.
The \Ha\ map reveals four structures: 1) the star formation ring, 2) the spiral arms, 3) biconical outflows, and 4) a doughnut shape.
First, the star formation ring is clearly detected in the \Ha\ map, although the shape is not a perfect ring.
The \Ha\ flux is stronger in the half ring in NW, while the other half ring in SE is relatively weak. This may be caused by the contamination or overlap of foreground AGN gas outflows to the line-of-sight (see Section 4.2.3). 
In the NW ring, we find an interesting region with prominent \Ha\ emission, which is marked as Region A in the left panel of Figure~3.
Region A is likely to be an intersection of the star formation ring and AGN gas outflows, which will be investigated in the following sections.  
Second, we detect weak \Ha\ emission along the spiral arms, particularly in the southern arm. These features indicate star formation activity in the spiral arms. 
Third, as previously reported \citep[see e.g.,][]{Durre2018,Durre2019}, \OIII\ flux map shows a conical shape from the center to 1-2 kpc scales in SE direction, indicating gas outflows. Finally, a doughnut shape is detected in the outflow region at the distance of $\sim$4\arcsec\ from the center. 
The doughnut shape could be interpreted as representing the hollow cone structure as suggested by previous outflow studies \citep[e.g.,][]{Fischer2010}. 
The \Ha\ and \OIII\ flux maps show similar trend except for the nuclear star formation ring and spiral arms, where star formation is expected, 
indicating that there are various ionizing sources in the FOV and this issue will be described in the Section 4.2.3.

\subsubsection{Kinematics}\label{astrometry correction}

The velocity maps of \Ha\ and \OIII\ with respect to the systemic velocity of NGC 5728 are 
presented in the upper panels of Figure~4, showing a rotation pattern due to the host galaxy gravitational potential,
while there is additional non-gravitational components in NW-to-SE direction. At the very center ($<$ 1 kpc), where the biconical outflows are detected, the location of the two clumps with the highest blueshift/redshift is misaligned with the larger scale gravitational motion (N-S direction). 
Also, we detect high velocity structures close to, but outside of the spiral arms. The blueshifted region in NW and the redshifted region in SE are clearly present in the \OIII\ velocity map, indicating the presence of non-gravitational motion, i.e., outflows, in $\sim$2 kpc scale.

{ Gas inflows are often detected along spiral arms, as gas emission lines are blueshifted in the far-side while the near-side presents redshifted emission lines \citep[e.g.,][]{Riffel2008,Riffel2013,Diniz2015,Luo2016}. In contrast, NGC 5728 shows an opposite trend, as \Ha\ and \OIII\ are redshifted in the far side (SE), while in the near side (NW) 
\Ha\ and \OIII\ show blueshift. Thus, the gas in the far-side (near-side) is receding (approaching), suggesting outflows rather than inflows.  }

To separate the non-gravitational motion, we construct the relative velocity maps by subtracting stellar velocity from gas velocity in each spaxel (lower panels of Figure~4). The relative motion of the ionized gas clearly shows the biconical outflows at the central region within the location of the star formation ring (hearafter inner region, see Figure~3). They are composed of a pair of receding (redshifted) and approaching (blueshifted) parts. 
We find that the gas outflows are not confined in the 1 kpc scale. Rather, gas outflows extend to 2 kpc scales slightly beyond the location of the spiral arms (hearafter outer region). 
The inner region shows the maximum \Ha\ velocity of 250 and -140 \kms\ in the receding and approaching cones, respectively, while 
the maximum velocity of the outer region is 160 and -180 \kms\ in the receding and approaching cones, respectively. 
The relative velocity maps of \Ha\ and \OIII\ show qualitatively similar morphology, however, \OIII\ generally shows more negative velocity in the outer region. The maximum velocities of \OIII\ in the inner and outer regions are, respectively, 320 (-140) and 90 (-220) \kms\ for the receding (approaching) cone.

The velocity dispersions of \Ha\ and \OIII\ tend to be higher in regions where the gas outflows are detected (i.e., inner and outer regions) compared to the rest of our FOV (see Figure~5). In the gas outflow regions, the observed velocity dispersion is the highest at the central part ($\sim$300 and $\sim$360 \kms\ for \Ha\ and \OIII) and gradually decreases as a function of distance from the center (down to $\sim$100\kms\ in \OIII). 
The high velocity dispersion detected in the gas outflow region suggests that the ionized gas, in particular \OIII, is under strong influence of AGN. 
At the central part, we find an interesting morphology with very high velocity dispersions (i.e., $>$300 \kms) along the perpendicular direction to the orientation of the 
gas outflows { (see also Figure 7 of \citealt[][]{Durre2019}), which also has been reported in other AGNs \citep[e.g.,][]{Riffel2014,Lena2015, Couto2017}. This feature can be interpreted as an equatorial outflow \citep{Riffel2014}, while seeing effect is likely to be the case (See Section 4.4 for more discussion).}

{ In order to investigate the kinematical relation between ionized gas and stars, we compare their V$_{\rm rms}$ (i.e., $\sqrt{V^{2}+\sigma^{2}}$) \citep[e.g.,][]{Cheung2016}. As shown in Figure~6, the V$_{\rm rms}$ ratio between ionized gas (\Ha\ and \OIII) and stars is high (i.e., $\sim$2) at the region, where gas outflows are detected. This result confirms the non-gravitational kinematics (i.e., gas outflows) in those regions, including the central part where a compact stellar disk is present. On the other hand, the V$_{\rm rms}$ ratio is generally close to unity in the rest of the FOV, indicating that gas follows the gravitational potential (see also the middle panel in Figure~2.}

\begin{figure*}{}
\includegraphics[width = 0.9\textwidth]{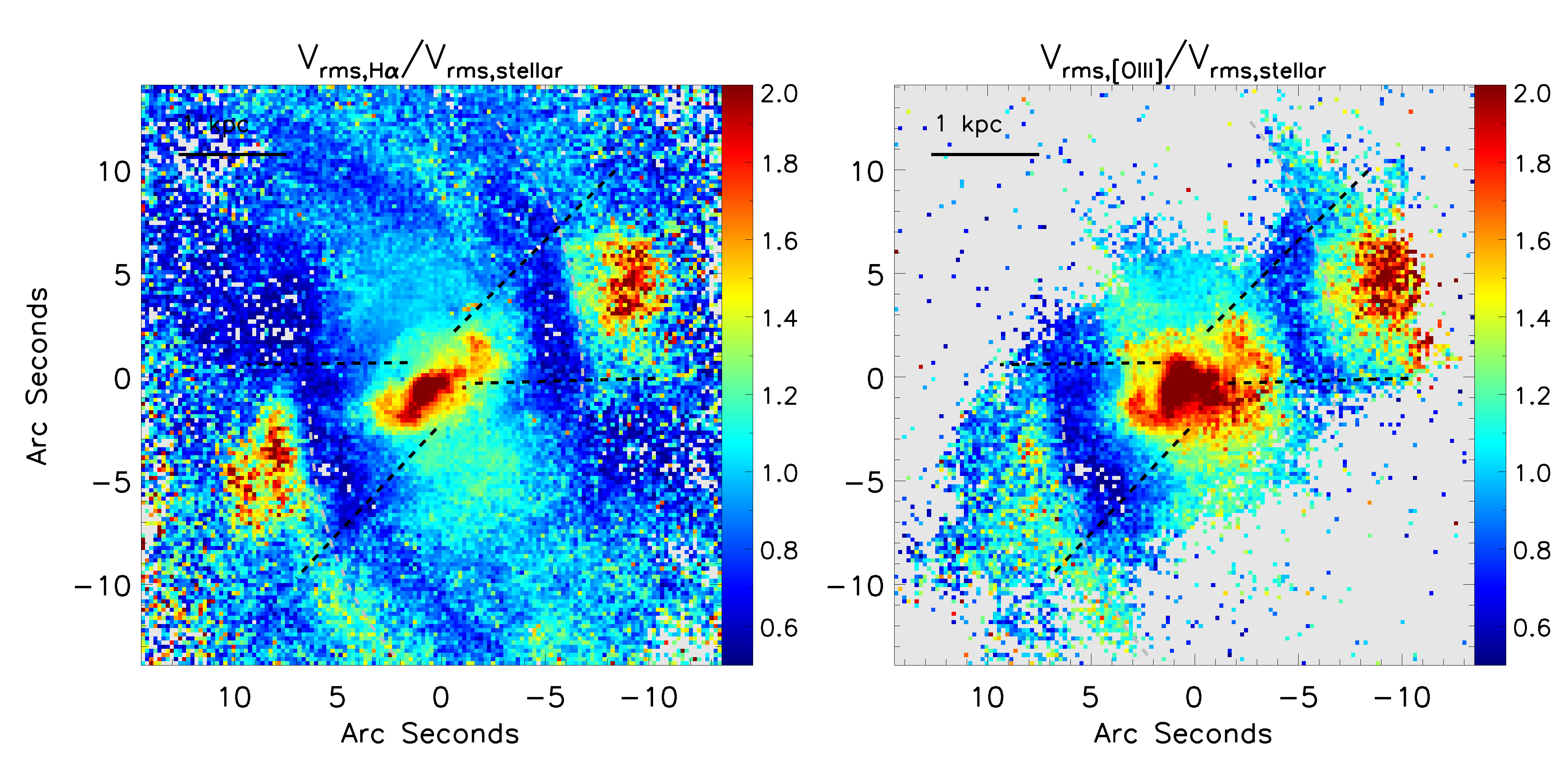}
\caption{
V$_{\rm rms}$ maps of \Ha\ (left) and \OIII\ (right). Spiral arms and biconical outflows are denoted with gray dashed lines and black dashed lines, respectively, as presented in Figure~3. Gray and black dashed lines are same as in Figure~3.
\label{fig:allspec1}}
\end{figure*}

Using the \OIII\ velocity dispersion, we estimate the size of the AGN gas outflows by adopting the method of \cite{Kang2018}. 
The outflow size is defined at the radius where \OIII\ velocity dispersion becomes comparable to stellar velocity dispersion. 
To be consistent with \cite{Kang2018}, we adopt stellar velocity dispersion of 160 \kms, that is measured from the integrated spectrum within 3\arcsec\ diameter at the center. Our estimated outflow size is $\sim600$ pc and the luminosity of \OIII\ within the outflow size is $8.51\times 10^{39}$ erg/s. These measurements are consistent with the outflow size-luminosity relation of \cite{Kang2018} within the scatter of 0.1 dex. This result indicates that if NGC 5728 is at a large distance, outflows will be mainly detected  from the region where gas velocity dispersion is very high, resulting in a relatively small outflow size. In contrast, we are able to detect gas outflows at much larger scales with high spatial resolution and high sensitivity, although the gas outflows are relatively weak. \\

 \begin{figure*}{}
\includegraphics[width = 0.98\textwidth]{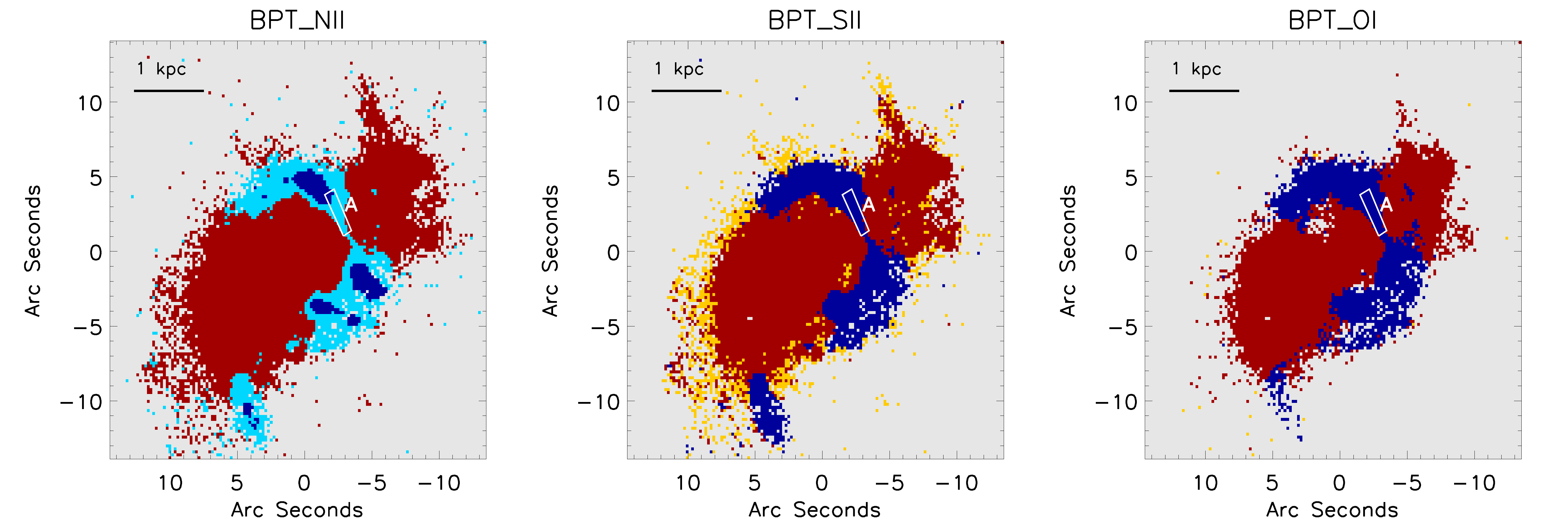}
\caption{
BPT morphology maps with three diagnosis (\NII, \SII,\ and \OI).
Red, blue, cyan, and yellow represent AGN, star forming region, composite, and LINER, respectively. 
 In each BPT map, Region A is marked with a white box. 
\\
\label{fig:allspec1}}
\end{figure*}

\begin{figure*}{}
\includegraphics[width = 0.98\textwidth]{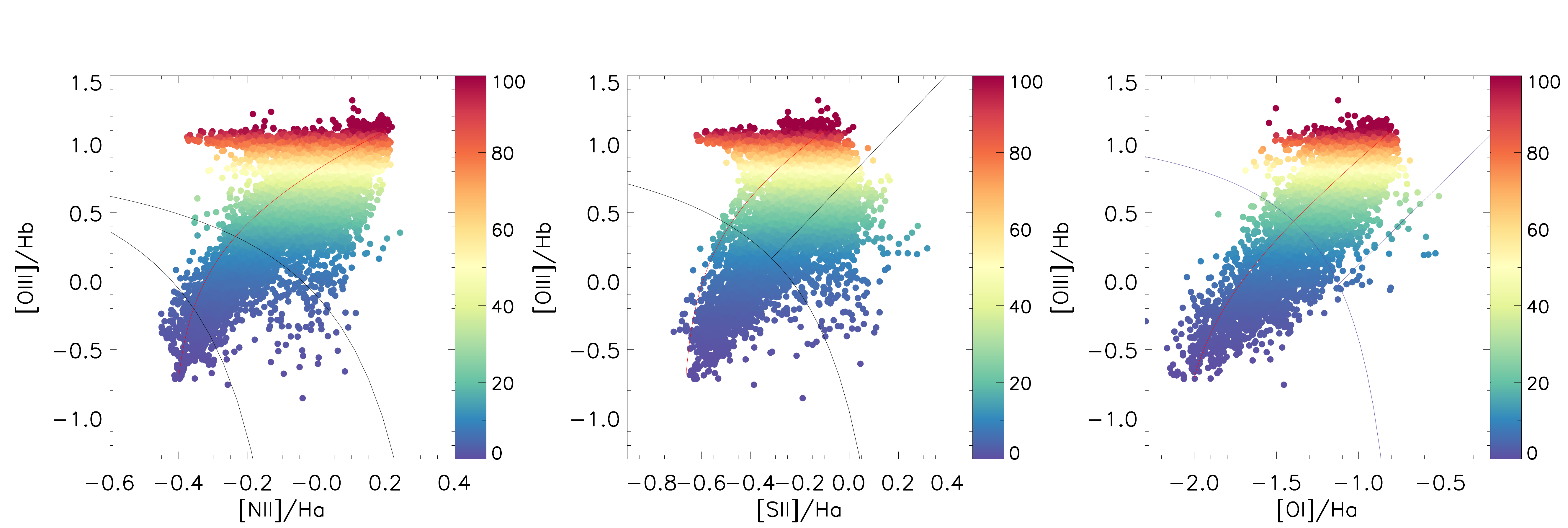}
\caption{
BPT diagrams with three diagnosis (\NII, \SII,\ and \OI). Red lines denotes the mixing sequence between the two basis points 
for star forming region and AGN region. Color presents AGN fraction from 0 to 100\%\ along the mixing sequence.\\
\label{fig:allspec1}}
\end{figure*}

\subsubsection{photoionization: AGN vs. star formation}\label{astrometry correction}
To investigate ionizing sources across the FOV, we investigate line flux ratios and identify
their ionizing sources in each spaxel, using the BPT diagrams with three diagnoses of \NII\, \SII\, and \OI. 
As shown in Figure~7, our BPT classification result well separates the AGN gas outflows and the star
formation ring as AGN and star forming region.
Also, the BPT diagrams indicate a clear mixture of photoionization due to star formation and AGN in the central part of NGC 5728 (see Figure~8).
We note that our BPT classification result is consistent with that presented in the previous works \citep[][]{Davies2016,Durre2018}.  

We separate ionizing sources using the emission line flux ratios in the BPT diagrams \citep[e.g.,][]{Kauffmann2009,Davies2016}, as similarly performed by \cite{Davies2016}. 
First, we determine two basis points, respectively, representing pure star formation and pure AGN in the BPT diagrams and draw a line between the two basis points as the mixing sequence (see red lines in Figure~8). Thus, the mixing sequence represents the AGN fraction from 0\% to 100\% between the two basis points. Note that the mixing sequence is curved in the log scale BPT diagrams, but it is defined as a line in the linear scale flux ratio diagrams. Then, for a given location in the BPT diagrams, we adopt the AGN fraction of the closest point in the mixing sequence. In this way we determine the AGN fraction of each location in the BPT diagram as presented in Figure 7. Note that we 
calculate the minimum distance to the mixing sequence using the linear scale (instead of log scale) BPT diagrams, following \citet{Davies2016}. 
The AGN fraction is color-coded in Figure~8. While the AGN fraction is determined independently using each BPT diagram, 
we find only marginal difference of the AGN fraction among them.
Thus, we adopt the AGN fraction estimated from the BPT diagram based on \NII. 
Comparing Figure~7 and 8, the AGN fraction is low (<$\sim$10\%) in star forming region while 
it goes up to 50-100\%\ in AGN region. 

Using the determined AGN fraction in each spaxel, we separate the contribution from star formation to \Ha\ emission, determine star-forming \Ha\ luminosity,
and calculate SFR based on Eq. 4 of \cite{Murphy2011}:
\begin{equation}
SFR(M_{\odot}\ yr^{-1}) = 5.37 \times 10^{42}\ L_{\rm H\alpha}\ \rm (erg/s).
\end{equation}
For the luminosity calculation, \Ha\ flux is corrected for dust extinction using Eq. A10 of \cite{Vogt2013} with the Balmer decrement (i.e., \Ha/\Hb) of 
2.86 and $R^{A}_{V}$ of 4.5 \citep{Fischera2005}.

The \Ha\ luminosity maps are presented in Figure~10, after separating AGN and star formation contribution to \Ha, along with the SFR map.

The AGN outflows and the star formation ring are well separated as consistent with those in \cite{Davies2016}.  
Moreover, we find that clumpy structures in the star formation ring become more prominent in the SFR map (see also Figure~3).
To compare with Region A, we additionally select three regions (i.e., Region B, C, and D) whose SFR is 
distinctively high in the star formation ring.
We arbitrary determine their size and estimate their SFR (Table~1).
We note that there is another region with high SFR 
in the south of Region C. 
However, AGN is dominant as the BPT classification indicates. Thus, we do not consider this region in the comparison
with Region A as a conservative approach. 

As shown 
in Figure~9, these regions show small AGN contributions (10$\pm$6, 5$\pm$6, 3$\pm$1, and 2$\pm$2 \% for 
Region A, B, C, and D, respectively), suggesting that most \Ha\ emission 
is coming from 
star formation.
Interestingly, we find that the SFR in Region A is the second highest among those of the four regions. Note that this trend 
does not change even if we consider three sigma uncertainties (up to $\sim$28\%) of the AGN fraction.

Similar to the \Ha\ flux map (see Figure~3), the SFR map (right panel of Figure~10) 
does not reveal the other half ring in SE direction even after the dust extinction correction, indicating that 
the dust correction was not successful. Since the fluxes of \Ha\ and \Hb\ in the SE half ring are largely dominated by 
AGN ($\sim$50-100\%), the dust extinction correction in that region is heavily weighted to that of the AGN cone, not to the 
star formation ring \citep[see also][]{Durre2018}. However, this issue is not relevant to the four regions defined in NW, because the 
AGN fractions in those regions are very low (<10\%).

Note that our AGN-star formation separation is consistent with that of \cite{Davies2016}, while it is somewhat different from 
that of \cite{Durre2018}. This is due to different approaches in the ionizing source separation. \cite{Durre2018} separated the ionizing 
sources with a logarithmic superposition with two basis points in the BPT diagrams, which is different from the linear superposition used in this work and
the work by \cite{Davies2016}. As shown in Figure~18 of \cite{Durre2018}, the AGN fraction is $\sim$40\%\ in the star formation ring, which is much larger
than our estimate ($\sim$10\%). However, even if we adopt the AGN fraction of $\sim$40\%, the SFR in Region A is still 
comparable to that in Region C and D.
\\

 \begin{figure}{}
\includegraphics[width = 0.44\textwidth]{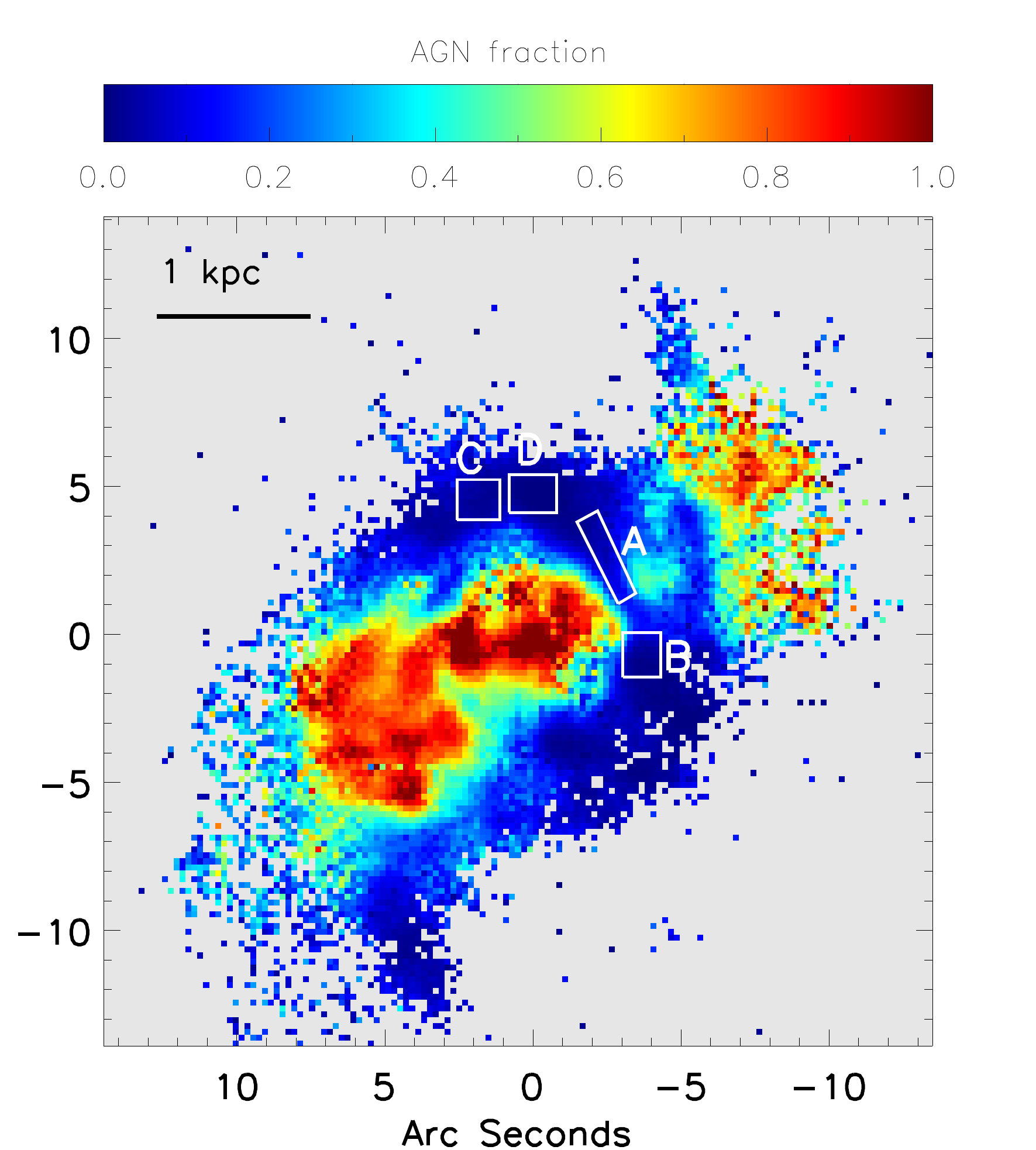}
\caption{AGN fraction map. Color represents the AGN contribution, ranging from 0 to 100\%. White boxes present Region A-D, which will be described in Section 4.2.3.
\label{fig:allspec1}}
\end{figure} 

 \begin{figure*}{}
\includegraphics[width = 0.99\textwidth]{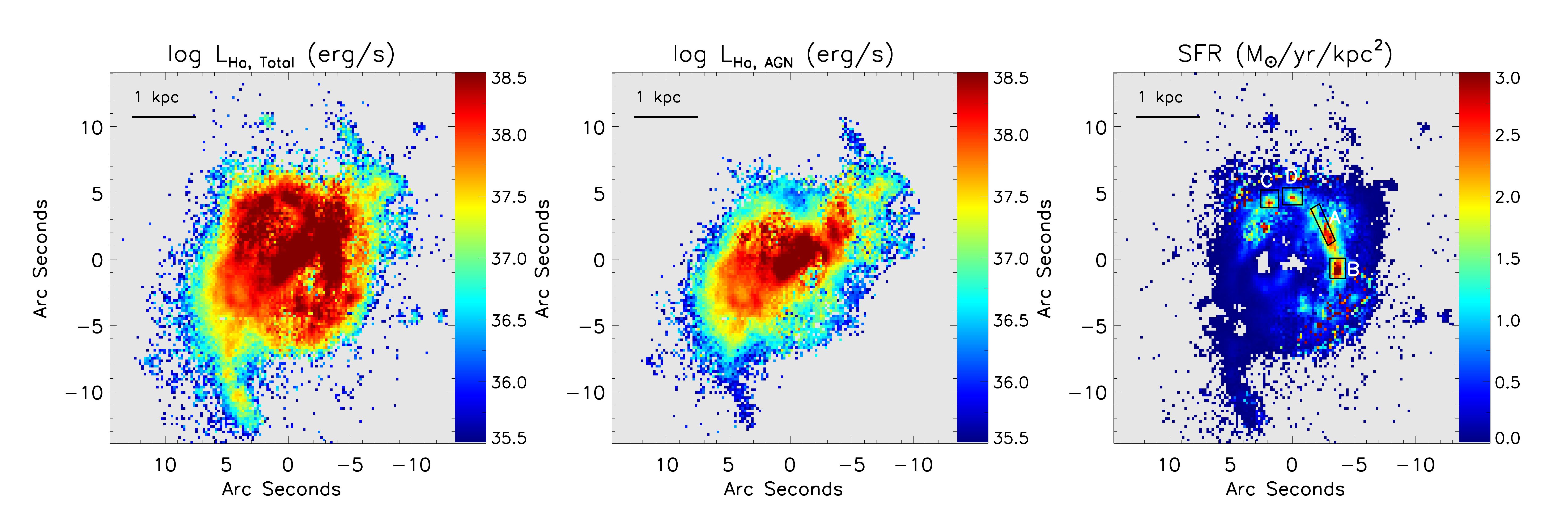}
\caption{
\Ha\ luminosity maps using the total \Ha\ line (left) and contribution from AGN (center). The SFR derived from the \Ha\ line after subtracting AGN contribution is
shown in the right panel. Region A-D with high SFR are marked with black boxes. \\
\label{fig:allspec1}}
\end{figure*} 

 \begin{figure*}{}
 \centering
\includegraphics[width = 0.92\textwidth]{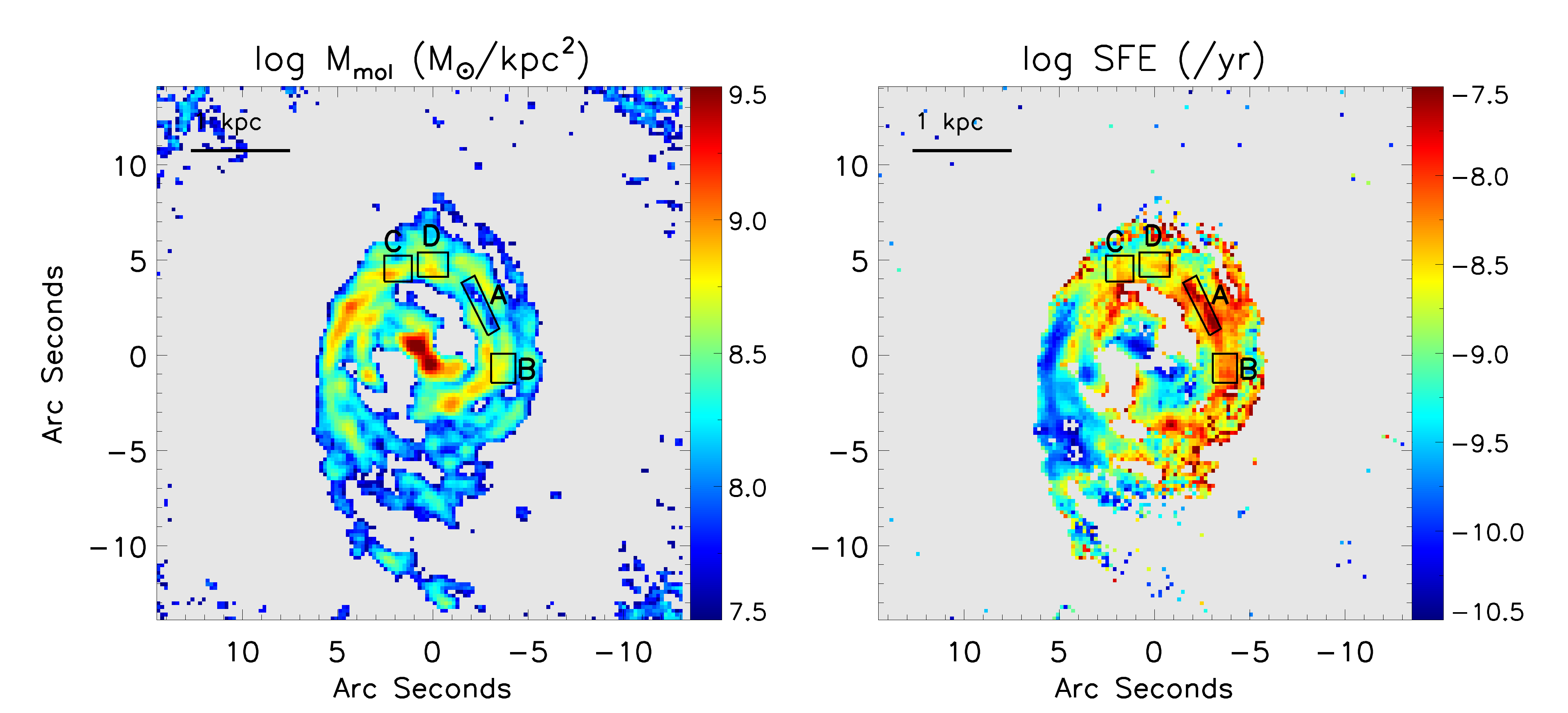}
\caption{
Left panel: molecular gas mass ($M_{\rm Mol}$) map derived from the CO (2-1) intensity map (left), showing various structures 
(i.e., a circumnuclear disk, streams, and clumps). 
Right panel: star formation efficiency (i.e., SFR/M$_{\rm Mol}$). Region A-D are marked with black boxes in each panel.\\
\label{fig:allspec1}}
\end{figure*}

\begin{deluxetable*}{ccccccc} 
\tablewidth{0pt}
\tablecolumns{7}
\tabletypesize{\scriptsize}
\tablecaption{SFR and SFE}
\tablehead{
\colhead{} &
\colhead{SFR\_avg} &
\colhead{SFR\_med} &
\colhead{SFE\_avg} &
\colhead{SFE\_med} &
\colhead{$\rm T_{dep}$\_avg} &
\colhead{$\rm T_{dep}$\_med} 
\\
&
\colhead{($\rm M_{\odot}/yr/kpc^2$)} &
\colhead{($\rm M_{\odot}/yr/kpc^2$)} &
\colhead{($\rm yr^{-1}$)} &
\colhead{($\rm yr^{-1}$)} &
\colhead{($\rm Gyr$)} &
\colhead{($\rm Gyr$)} 
\\
\colhead{(1)} &
\colhead{(2)} &
\colhead{(3)} &
\colhead{(4)} &
\colhead{(5)} &
\colhead{(6)} &
\colhead{(7)} 
}
\startdata
Region A & $1.82$ & $1.70$ & 1.62$\times \rm 10^{-8}$ &  1.15$\times \rm 10^{-8}$ & 0.062 & 0.087 \\
\hline
Region B & $2.15$ & $1.98$ & 5.25$\times \rm 10^{-9}$ &  5.33$\times \rm 10^{-9}$ & 0.190 & 0.188 \\
\hline
Region C & $1.20$ & $1.13$ & 3.14$\times \rm 10^{-9}$ &  2.84$\times \rm 10^{-9}$ & 0.318 & 0.352 \\
\hline
Region D & $1.04$ & $0.90$ & 2.47$\times \rm 10^{-9}$ &  2.34$\times \rm 10^{-9}$ & 0.405 & 0.427
\enddata
\label{table:prop}
\tablecomments{Averaged or median value of SFR, SFE, and depletion time scale
for the four regions marked in Figure~10 and 10. }
\end{deluxetable*}

\subsection{Molecular gas} 
To investigate the molecular gas and its relation with SFR, we present the distribution of the molecular gas mass 
($M_{\rm Mol}$) and star formation efficiency (SFE=SFR/$M_{\rm Mol}$) 
in Figure~11. In each panel, Region A-D with high SFR are marked.
The structure of the molecular gas generally follows that of the SFR map (Figure~10) with showing the star formation ring
and the spiral arms. However, the molecular gas distribution also shows distinct features 
(i.e., a circumnuclear disk, non-detection along the outflow orientation, streams, and clumps). 

First of all, the distribution of the molecular gas is strongly concentrated at the central part, 
which may indicate a circumnuclear disk with the radius of 1\arcsec\ corresponding to 200pc, 
similar to e.g., NGC 1068 \citep{Garcia-Burillo2014}. { As shown in the map of stellar velocity (the middle panel of Figure~2), 
the centrally fast rotating motion of the stars may be related to the circumnuclear disk.}
Remarkably, the position angle of the circumnuclear disk is $\sim$40\arcdeg, which is nearly perpendicular to the orientation of the gas
outflow (i.e., -53\arcdeg). This may indicate that the dust torus of AGN may be confined in the circumnuclear disk although
it is not resolved at the given resolution of our ALMA data. 

Secondly, we do not detect CO along the outflow orientation, which is a different trend from that of the ionized gas.
This non-detection can be due to the excitation or dissociation of CO by X-ray photons (or radio jets) from the AGN. 
This issue will be discussed in the next section. 

Thirdly, we detect CO gas streams along the star formation ring at the position angle of $\sim$45\arcdeg and -135\arcdeg (i.e., Region B). These regions may be related to contact points between the star formation ring and large scale structures of the host galaxy.
Due to the large amount of inflowing materials along the spiral arms, molecular gas can be accumulated hence star 
formation can be active near the contact points \citep[e.g.,][]{Boker2008}. Actually, in Region B, we detect the 
highest star formation compared to the other regions in the star formation ring (see Table~1). 

Finally, clumpy regions are detected (e.g., Region C and D). 
With high molecular mass as well as SFR, these regions may be typical star forming regions. 

Contrary to Region B, C, and D, CO emission is barely 
detected at Region A. Consequently, the observed SFE (depletion time scale) 
in Region A is $\sim$3-5 times higher (shorter) than that of other regions (see Table~1).
If we consider Region B, C, and D as typical star forming 
regions without AGN feedback, while region A could have a different mechanism for 
star formation (i.e., AGN feedback). We will discuss this issue in the next section.\\

\subsection{Bicone model of AGN gas outflow} 

We constrain the geometry of the gas outflows using a bicone model. Several efforts have been made to 
model the biconical outflows \citep[e.g.,][]{Fischer2013,Bae2016}. With various structural parameters (i.e., inclination and outer opening angle),
the observed kinematics of the gas outflows are reproduced. However, the previous models did not consider seeing effect 
which can severely change the observed morphology of the gas outflows in the projected plane. 

To apply the seeing effect, we build a three dimensional bicone model by including a point spread function in the model of \cite{Bae2016}. The updated model requires 11 free parameters (i.e., 7 parameters for the bicone geometry, 3 parameters for a dust plane for extinction effect, and one 
parameter for the seeing size). A detailed description for the updated model with the seeing effect will be presented in the future (Shin et al. in preparation). 

Through various tests, we determine the best parameters of the bicone model, which represents the observation with the minimum chi-square value,
under three specific conditions as described below. 
First, we focus on the central region ($\sim10\arcsec \times 10\arcsec$), where the AGN fraction is dominant, to minimize the contamination from host galaxy 
since this model only accounts for non-gravitational effect due to AGN outflows. Note that even though we separate the ionizing sources (i.e., AGN and star formation), 
the kinematics are not separated in this work. Second, to reduce the number of geometry parameters 
in the modeling, we adopt the seeing size as 0.66\arcsec\ from the MUSE observation, and constrain the geometry of the large scale dust plane from the observed properties of the star forming ring, i.e., the position angle of the major axis of the ring: 20\arcdeg, and the inclination of the minor axis: 50\arcdeg\ based on the previous work by \citet{Son2009a}. Lastly, we compare one dimensional radial profiles of the flux, velocity, and velocity dispersion of the bicone model along the AGN gas outflow direction 
(see a thick black line in Figure~3-5) with the observation of \OIII, in order to determine the best physical parameters of the gas outflows. 
Note that, \OIII\ velocity dispersion suddenly decreases in 2-3\arcsec in the approaching cone (shaded area in Figure~12), of which the origin is not clear.
For the comparison, we masked out this region. The dip feature may be caused by the contamination from star formation ($\sim$20\%) or the dust obscuration of the gas outflows.  An interaction between host galaxy and AGN gas outflows is also a potential explanation of the dip feature \citep[e.g.,][]{Fischer2017}.


Figure~12 presents the radial profiles of the flux, velocity, and velocity dispersion, reproduced from the best bicone model,
compared to the observed profiles, as a function of distance from the position of X-ray source \citep[RA=14:42:23.88 and Dec=-17:15:11.25][]{Evans2010}.
While the reproduced radial profile of \OIII\ kinematics are not perfectly matching the observations,
the model reasonably well explains the measured velocity and velocity dispersion of \OIII. 
The best bicone model is constrained with an inclination of 20\arcdeg\ and an outer opening angle of $\pm$28\arcdeg. 
This results suggest that the gas outflows cover the relatively wide inclination angle from -8 to +48\arcdeg, and encounters the stellar disk including the star formation ring and the spiral arms, of which the inclination angle of the minor axis is $\sim$50$\arcdeg$ \citep{Son2009a}. 

Also, the smaller inclination of the bicone than that of the star formation ring confirms that the approaching cone in NW direction is behind the star formation ring while the receding cone in SE direction is in front of the ring to the line-of-sight. 

We present the simulated maps of flux, velocity, and velocity dispersion based on the best bicone model (Figure~13). 
Even though our model does not fully represent the observations due to various limitations in the model and various issues in the observations 
(i.e., star formation contribution and non-uniform dust extinction), the simulated maps qualitatively represent the flux distribution and kinematics of \OIII. 

One striking result in our two dimensional model is the elongated feature with the highest velocity dispersion, which is orientated perpendicular to 
the direction of the gas outflow. This feature is clearly detected in the velocity dispersion map of \OIII\ (Figure~5), suggesting that the seeing effect { can be} responsible for this structure. Note that this perpendicular feature can not be reproduced without the seeing effect in our previous model \citep[see Figure~3 of][]{Bae2016}, indicating the importance of the seeing effect. 
{ Since the effect of the overlap between approaching and receding cones is the strongest at the central region, velocity dispersion
is naturally expected to be very high. The seeing effect artificially increases the front of the overlap between approaching and receding cones in the perpendicular direction,
while in the outflow direction, the seeing effect is much weaker due the the contribution from the outer region. Thus, the perpendicular shape of the highest velocity 
dispersion is produced in our bicone model. 
In our simulation, the size of the perpendicular feature is 2.5\arcsec, which is smaller than that of the observed feature ($\sim$4.4\arcsec). 
Nonetheless, we qualitatively confirm the possibility that the perpendicular feature is due to the seeing effect.

Equatorial outflow is another possibility. For example, \cite{Riffel2014} discussed equatorial outflows aligned with an extended radio emission in NGC 5929. 
To investigate the possibility of equatorial outflows, we examine the 20 cm radio image of NGC 5728 \citep{Schommer1988}. While there are weak radio emissions in the central region, 
we find no clear evidence of an extended radio emission along the perpendicular direction with respect to the outflow direction. 
}

We note that \cite{Durre2019} also constrained the inclination and the outer angle of the bicone and discussed the possible 
interaction between host galaxy-AGN outflow. Even though their results are generally consistent with ours, 
the values of the inclination (47.6\arcdeg) and outer angle (71\arcdeg) are quantitatively different. 
If their inclination value is adopted, it is difficult to expect the dust obscuration of the approaching cone in NW direction by the star formation ring or dusty stellar disk,
since the inclination of the cone and the disk is very similar. 
The reason for the discrepancy seems to be due to their specific analytic model of gas outflows in \cite{Durre2019}.
For example, they constrained the geometry (i.e., inclination and opening angle) with the assumption that 
the velocities of [\FeII] 1.644 $\mu$m, which were measured from each Gaussian component in the best-fit model, 
represent the front and back velocities of the hollow cone (see Eq. 14 and 15 of \citealt{Durre2019}). 
However, 
the front and back velocities measured from the two Gaussian components in the emission line, do not represent the same distance from the center due to the projection effect, which varies depending on the inclination and opening angle. 
Thus, more detailed constraints are needed to fully understand the geometry of the outflows.

 \begin{figure}{}
\centering
\includegraphics[width = 0.44\textwidth]{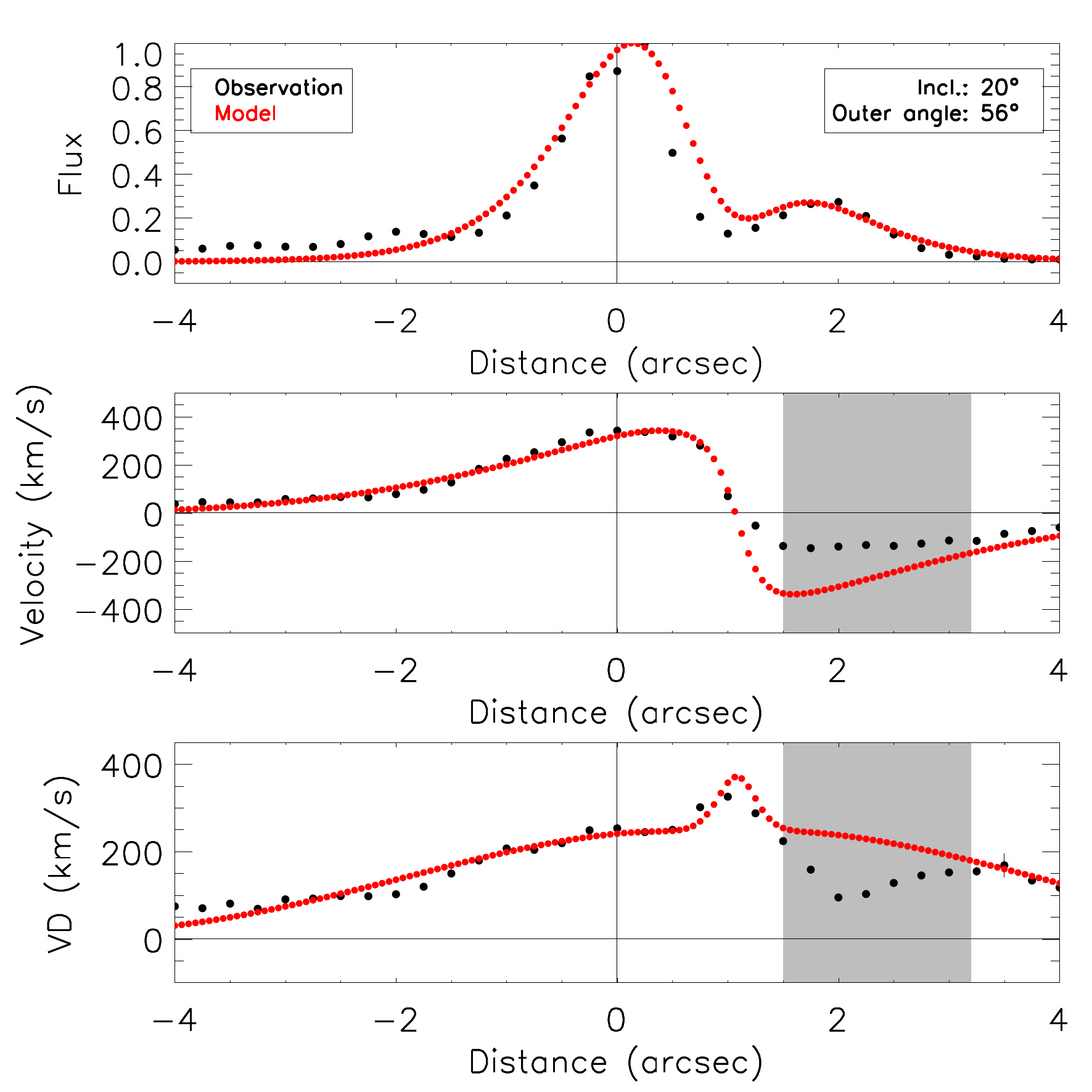}
\caption{
Radial profiles of the measured flux, velocity, velocity dispersion of \OIII\ (black points) along the AGN gas outflow direction (see the pseudo-slit in Figure~3). 
The radial distance in the projected plane is measured from the position of X-ray source.
Red points represent the measurements from the best bicone model with an inclination of 20$\arcdeg$ and the outer angle of $\pm$28$\arcdeg$. 
The region from 1.5 to 3.2\arcsec\ is masked out for the comparison between observations and model predictions. 
\label{fig:allspec1}}
\end{figure} 

 \begin{figure*}{}
 \centering
\includegraphics[width = 0.96\textwidth]{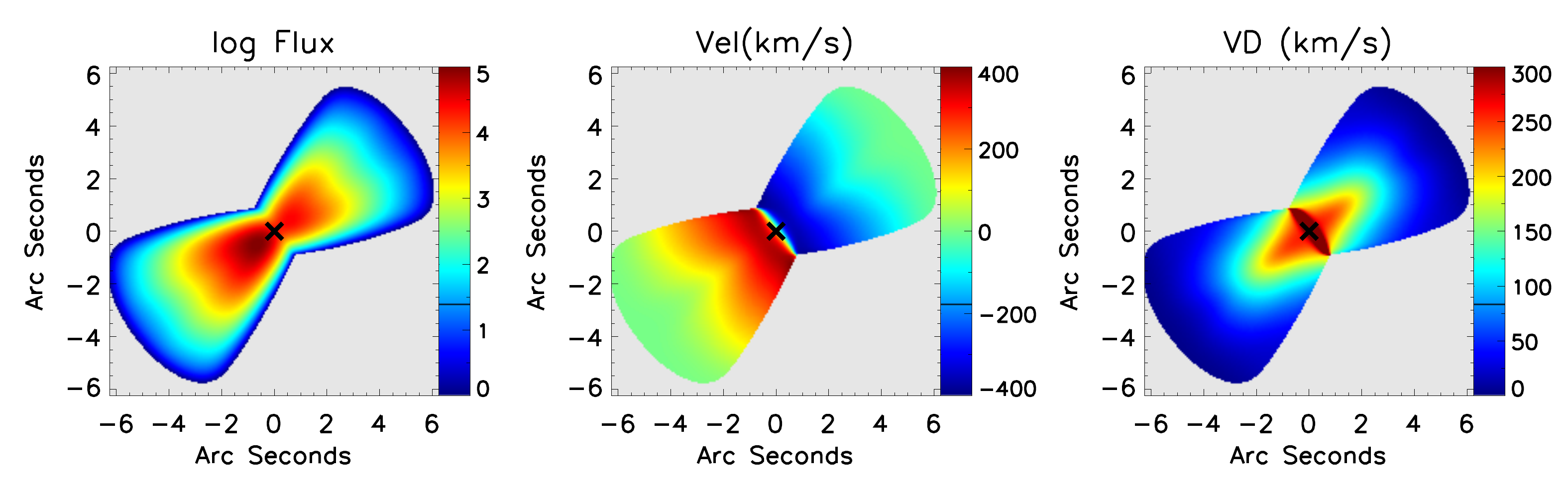}
\caption{
Simulated maps of flux, velocity, and velocity dispersion using the biconical outflow model.
The center of the outflows is marked with cross. The model parameters are same as in Figure~12. 
\\
\label{fig:allspec1}}
\end{figure*} 

 \begin{figure}{}
 \centering
\includegraphics[width = 0.44\textwidth]{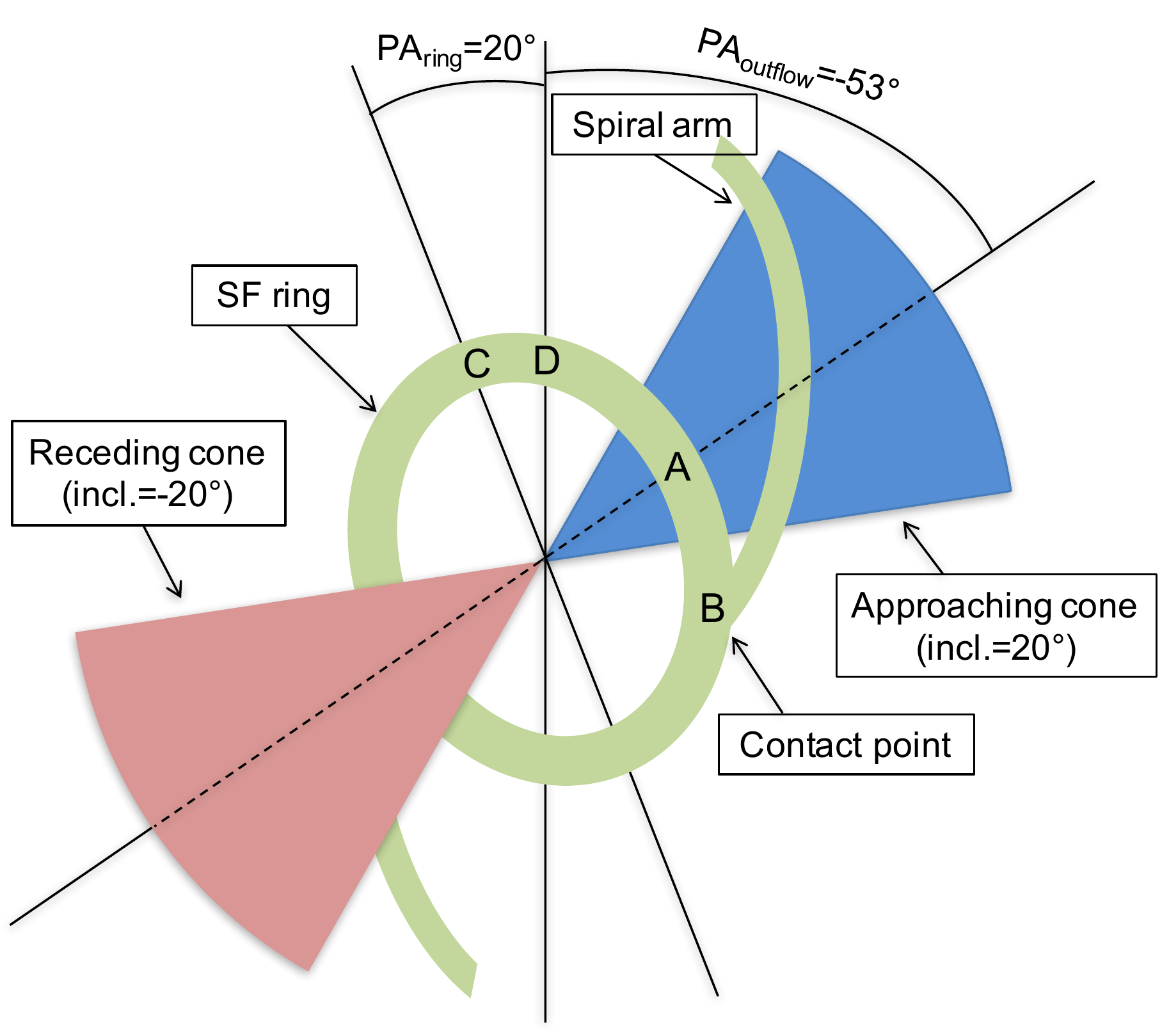}
\caption{
Schematic view of the central part of NGC 5728. 
The PA of the biconical outflows is -53\arcdeg\ and the inclination angle is 20\arcdeg\ with an opening angle of $\pm$28\arcdeg,
while the PA of the major axis of the ring is 20\arcdeg and the inclination angle of the minor axis of the ring is 50\arcdeg. The encountering region between the gas outflows and the ring is denoted with A, while the highly star-forming regions in the ring are dented with C, and D. 
B represents the contact point between the ring and the northern spiral arm. 
\label{fig:allspec1}}
\end{figure} 

Based on the constraints in this study, we construct a schematic model, consisting of the star formation ring, 
the spiral arms and the biconical gas outflows (see Figure~14). As expected from the MUSE observation, the geometry 
of the model well represents that the approaching cone is obscured by the star formation ring, while the
receding cone is in front of the star formation ring. 

As a reference, we mark the four regions with high SFR (i.e., Region A-D). In particular, Region A is presented
as the interaction region between the AGN gas outflows and the star formation ring. Region B is marked at  
the contact point between the star formation ring and the northern spiral arm. 
Finally, we indicate Region C and D, which are regarded as typical star forming regions 
in the star formation ring. \\

\section{Discussion}\label{Discussion}

\subsection{Positive feedback}\label{astrometry correction}

We detect the high SFR in Region A, where the gas outflows encounter the
star formation ring at a $\sim$1 kpc projected distance from the center (see Figure~10). The star formation efficiency of Region A is higher than other regions (B, C, and D) in the ring by a factor of $\sim$3-5, suggesting that the AGN-driven outflows may have enhanced star formation in the intersecting area. 
The relative deficit of CO molecular gas in Region A also indicates that the triggering mechanism of star formation 
is different compared to other regions, indicating the positive role of the AGN-driven outflows. 

A possible explanation of the lack of CO gas is that the AGN-driven outflows triggered a burst of star formation, consuming a 
large amount of molecular gas. Another scenario is that CO is excited by X-ray photons from AGN 
\citep[e.g.,][]{vanderWerf2010}. Since X-ray emission is detected in the gas outflow regions as well as in 
Region A \citep[see e.g.,][]{Durre2018}, the marginal detection of the CO (2-1) emission can be explained if 
CO is mostly excited to higher states (e.g., CO (3-2)) by X-ray photons.
Similarly, the excitation or dissociation of CO by shock may cause the lack of the CO (2-1) emission \citep[e.g.,][]{Flower2010,Meijerink2013}. 
These scenarios of the excitation/dissociation of CO can be also applied to the central part of NGC 5728, 
where the CO (2-1) emission is not detected along the outflow direction. Multi-phase CO observations are required to
verify these scenarios as the origin of the lack of the CO (2-1) emission in Region A. Nevertheless, all the proposed scenarios indicate the interaction between the AGN-driven outflows and the ISM in the star formation ring, supporting the positive feedback interpretation.  

We turn to the overall impact of the AGN outflows on star formation in NGC 5728. 
The estimated SFR in Region A is $\sim$ 0.2 M$_{\odot}$ yr$^{-1}$, which is only $\sim$10\%\ of the combined SFR ($\sim$ 2 M$_{\odot}$ yr$^{-1}$) in the $30 \arcsec \times 30 \arcsec$ FOV (i.e., $\rm 6 \ kpc \times 6\ kpc$). Regarding the total SFR in the entire galaxy, 
the contribution of the enhanced SFR due to the AGN outflows is much lower than 10\%. With this small 
contribution, it is difficult to claim a significant impact of the AGN outflows in NGC 5728.
While the overall effect of AGN feedback is limited, our results indicate that the AGN outflows 
may trigger star formation in high density regions (e.g., star formation ring or dust lane) 
as expected by several theoretical works \citep[e.g.,][]{Silk2013,Zubovas2017} 
and also reported by observational studies \citep{Cresci2015a,Cresci2015b,Carniani2016}.
\\

\subsection{Negative feedback}\label{astrometry correction}

We detect the two main regions of the gas outflows. While the inner region of the gas outflows at $<$ 1 kpc distance in the projected plane has been extensively discussed \citep[e.g.,][]{Wilson1993,Son2009a,Durre2018,Durre2019}, 
we newly find the outer region of the gas outflows at $\sim$2 kpc scale, by calculating relative velocity of ionized gas with respect to stellar velocity at each spaxel (see Figure~4). Interestingly, the location of the outer region is further out, compared to the location of the spiral arms. We interpret that inflowing gas along the spiral arms is swept out by the AGN outflows, presenting relatively high velocity and velocity dispersion in the outer region. Since gas supply is a key for star formation, this result implies that the star formation in the spiral arms is quenched due to the gas removal by AGN (i.e., negative feedback). Although the SFR in the outer region of the gas outflows is not well estimated due to the imperfect dust extinction correction, star formation activity seems much weaker than that in the star formation ring.

 \begin{figure}{}
 \centering
\includegraphics[width = 0.44\textwidth]{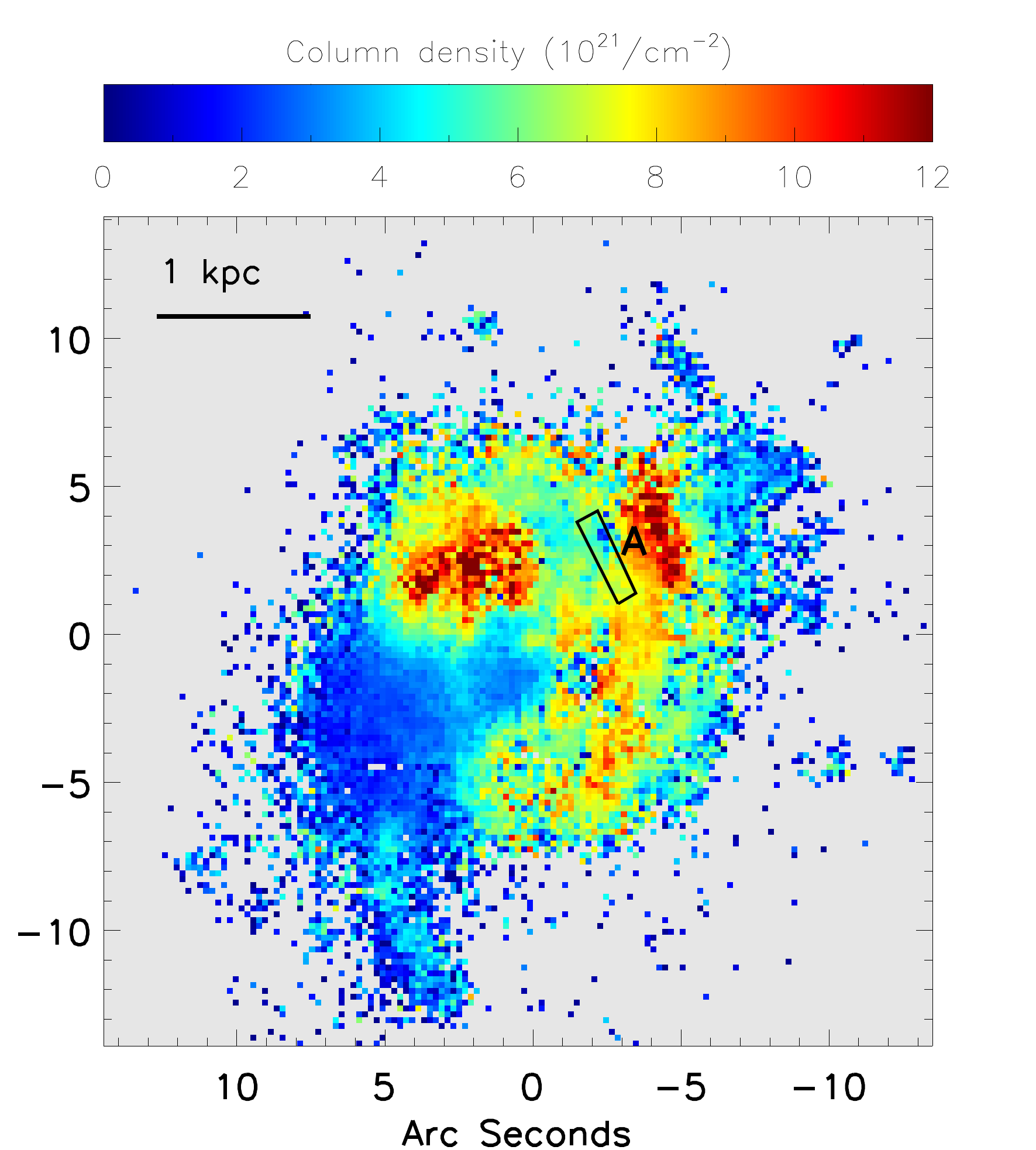}
\caption{
Hydrogen column density map calculated from the extinction. Region A is denoted with a black box.
\label{fig:allspec1}}
\end{figure} 

\subsection{Gas density and AGN luminosity}\label{astrometry correction}
The characteristics of AGN feedback have been explored with various gas density and AGN luminosity in theoretical studies. For example, \citet{Zubovas2017} showed that positive feedback was more likely detected in regions with high gas density while negative feedback was stronger as AGN luminosity increases.
To investigate the role of gas density in the context of AGN feedback, we estimate hydrogen column density (N$_{\rm H}$)
using the extinction magnitude \citep[e.g.,][]{Guver2009}, which is calculated with the Balmer decrement (see Figure~15).
The hydrogen column density in Region A ($\sim 6 \times 10^{21}\ \rm cm^{-2}$) significantly exceeds the critical density required for star formation \citep[$10^{21}\ \rm cm^{-2}$;][]{Clark2014}. On the other hand, in the outer region of the gas outflows in NW, where we interpret that gas is swept from the northern spiral arm without star formation activity, the detected N$_{\rm H}$ is $\sim 4 \times 10^{21}\ \rm cm^{-2}$, which is lower than that of Region A, but still higher than the critical density, suggesting star formation may be on-going. Thus, we find no clear evidence that gas density itself determines the nature of feedback (i.e., positive or negative). However, for the outer region of the gas outflows, it is limited to reliably determine the hydrogen column density due to the high contamination from AGN emission as the very high AGN fraction indicates ($\sim80\%$, see Figure~9). Thus, the dependence of AGN feedback on gas density needs to be investigated with further observations. 

Considering the effect of AGN luminosity, \cite{Zubovas2017} showed that negative feedback is effective 
when AGN luminosity is high ($L_{\rm Bol}>\ \sim4 \times 10^{46}$ erg/s), while a marginal effect in suppressing star formation was found 
in low AGN luminosity ($L_{\rm Bol}=1.3-2.6 \times 10^{46}$ erg/s). In the case of NGC 5728, the bolometric luminosity of AGN is $1.46\times\ 10^{44}\ \rm 
erg/s$ estimated from X-ray luminosity (\citealt{Durre2019}, see also \citealt{Davies2015}), which is far lower than that of AGNs explored by \cite{Zubovas2017}. 
Thus, considering the low luminosity of the AGN in NGC 5728, the overall negative feedback is not expected, which is consistent with our observations.  
Nevertheless, in small scales, AGN-driven outflows may suppress and enhance star formation, depending
on the physical properties, i.e., local density, as manifested in NGC 5728, although the overall impact of AGN outflows in the global star formation 
may not be significant.

\section{Summary}\label{summary}
In this work, we present the spatially resolved analysis of ionized and molecular gas in NGC 5728, focusing on the central 6 $\times$ 6 kpc scales, 
using the VLT/MUSE and ALMA data. We investigate AGN-driven outflows and their connection to star formation. The main results are summarized below.\\

1. We detect AGN-driven gas outflows out to $\sim$2 kpc from the center. 
While the inner region of the gas outflows at $<$ 1 kpc has been extensively studied, we newly present the outer region of the gas outflows based on the \OIII\ and \Ha\ gas velocity relative to that of stars in each spaxel. The inner and outer regions are disconnected by the star formation ring.
\\

2. We find that star formation activity is enhanced at the region where the AGN-driven gas outflows intervene the star formation ring, which can be interpreted as triggered by the AGN outflows. This positive feedback interpretation is supported by the deficit of the CO  (2-1) emission in that region compared to that of other regions in the ring. \\

3. The outer region of the gas outflows at $\sim$ 2 kpc scale is detected outside of the spiral arms, suggesting that the AGN outflows remove the inflowing gas from a large scale out of the spiral arms. We interpret this feature as an evidence of negative feedback.\\

4. Based on the three dimensional kinematical model, combined with the seeing effect, we reproduce the radial trend of gas velocity and velocity dispersion,
constraining the physical parameters of the biconnical gas outflows. The constraints of physical parameters further support that the AGN-driven outflows interact with ISM 
in the star formation ring. \\

5. Our results show the evidences of positive and negative feedback, while the overall impact of the AGN outflows on the total SFR is insignificant. For locally confined regions, gas density may play an important role in determining the characteristics of AGN feedback.\\

In this work, we find the complex nature of the AGN-driven outflows and their connection to the star formation in NGC 5728, along with the enhanced star formation as well as the gas removal. For better understanding of the role of AGNs in galaxy evolution, future works with 
a larger sample covering a large dynamic range in AGN luminosity are required.\\

\acknowledgements
{ We thank the anonymous referee for his/her valuable comments and suggestions that helped to improve the paper.}
This work was supported by the National Research Foundation of Korea grant funded by 
the Korea government (No. 2016R1A2B3011457 and No. 2017R1A5A1070354). 
Based on observations collected at the European Southern Observatory under ESO programme 097.B-0640(A).
This paper makes use of the following ALMA data: ADS/JAO.ALMA$\#$2015.1.00086.S. ALMA is a partnership of 
ESO (representing its member states), NSF (USA) and NINS (Japan), together with NRC (Canada) and NSC and 
ASIAA (Taiwan) and KASI (Republic of Korea), in cooperation with the Republic of Chile. The Joint ALMA Observatory 
is operated by ESO, AUI/NRAO and NAOJ.


\begin{thebibliography}{79}
\expandafter\ifx\csname natexlab\endcsname\relax\def\natexlab#1{#1}\fi

\bibitem[{{Bae} \& {Woo}(2014)}]{Bae2014}
{Bae}, H.-J., \& {Woo}, J.-H. 2014, \apj, 795, 30

\bibitem[{{Bae} \& {Woo}(2016)}]{Bae2016}
---. 2016, \apj, 828, 97

\bibitem[{{Bae} {et~al.}(2017){Bae}, {Woo}, {Karouzos}, {Gallo}, {Flohic},
  {Shen}, \& {Yoon}}]{Bae2017}
{Bae}, H.-J., {Woo}, J.-H., {Karouzos}, M., {et~al.} 2017, \apj, 837, 91

\bibitem[{{Barbosa} {et~al.}(2009){Barbosa}, {Storchi-Bergmann}, {Cid
  Fernandes}, {Winge}, \& {Schmitt}}]{Barbosa2009}
{Barbosa}, F.~K.~B., {Storchi-Bergmann}, T., {Cid Fernandes}, R., {Winge}, C.,
  \& {Schmitt}, H. 2009, \mnras, 396, 2

\bibitem[{{B{\"o}ker} {et~al.}(2008){B{\"o}ker}, {Falc{\'o}n-Barroso},
  {Schinnerer}, {Knapen}, \& {Ryder}}]{Boker2008}
{B{\"o}ker}, T., {Falc{\'o}n-Barroso}, J., {Schinnerer}, E., {Knapen}, J.~H.,
  \& {Ryder}, S. 2008, \aj, 135, 479

\bibitem[{{Bolatto} {et~al.}(2013){Bolatto}, {Wolfire}, \&
  {Leroy}}]{Bolatto2013}
{Bolatto}, A.~D., {Wolfire}, M., \& {Leroy}, A.~K. 2013, \araa, 51, 207

\bibitem[{{Braine} \& {Combes}(1992)}]{Braine1992}
{Braine}, J., \& {Combes}, F. 1992, \aap, 264, 433

\bibitem[{{Cano-D{\'{\i}}az} {et~al.}(2012){Cano-D{\'{\i}}az}, {Maiolino},
  {Marconi}, {Netzer}, {Shemmer}, \& {Cresci}}]{Cano-Diaz2012}
{Cano-D{\'{\i}}az}, M., {Maiolino}, R., {Marconi}, A., {et~al.} 2012, \aap,
  537, L8

\bibitem[{{Cappellari}(2017)}]{Cappellari2017}
{Cappellari}, M. 2017, \mnras, 466, 798

\bibitem[{{Carniani} {et~al.}(2016){Carniani}, {Marconi}, {Maiolino},
  {Balmaverde}, {Brusa}, {Cano-D{\'{\i}}az}, {Cicone}, {Comastri}, {Cresci},
  {Fiore}, {Feruglio}, {La Franca}, {Mainieri}, {Mannucci}, {Nagao}, {Netzer},
  {Piconcelli}, {Risaliti}, {Schneider}, \& {Shemmer}}]{Carniani2016}
{Carniani}, S., {Marconi}, A., {Maiolino}, R., {et~al.} 2016, \aap, 591, A28

\bibitem[{{Cheung} {et~al.}(2016){Cheung}, {Bundy}, {Cappellari}, {Peirani},
  {Rujopakarn}, {Westfall}, {Yan}, {Bershady}, {Greene}, {Heckman}, {Drory},
  {Law}, {Masters}, {Thomas}, {Wake}, {Weijmans}, {Rubin}, {Belfiore},
  {Vulcani}, {Chen}, {Zhang}, {Gelfand}, {Bizyaev}, {Roman-Lopes}, \&
  {Schneider}}]{Cheung2016}
{Cheung}, E., {Bundy}, K., {Cappellari}, M., {et~al.} 2016, \nat, 533, 504

\bibitem[{{Cicone} {et~al.}(2014){Cicone}, {Maiolino}, {Sturm},
  {Graci{\'a}-Carpio}, {Feruglio}, {Neri}, {Aalto}, {Davies}, {Fiore},
  {Fischer}, {Garc{\'{\i}}a-Burillo}, {Gonz{\'a}lez-Alfonso},
  {Hailey-Dunsheath}, {Piconcelli}, \& {Veilleux}}]{Cicone2014}
{Cicone}, C., {Maiolino}, R., {Sturm}, E., {et~al.} 2014, \aap, 562, A21

\bibitem[{{Clark} \& {Glover}(2014)}]{Clark2014}
{Clark}, P.~C., \& {Glover}, S.~C.~O. 2014, \mnras, 444, 2396

\bibitem[{{Couto} {et~al.}(2017){Couto}, {Storchi-Bergmann}, \&
  {Schnorr-M{\"u}ller}}]{Couto2017}
{Couto}, G.~S., {Storchi-Bergmann}, T., \& {Schnorr-M{\"u}ller}, A. 2017,
  \mnras, 469, 1573

\bibitem[{{Cresci} {et~al.}(2015{\natexlab{a}}){Cresci}, {Mainieri}, {Brusa},
  {Marconi}, {Perna}, {Mannucci}, {Piconcelli}, {Maiolino}, {Feruglio},
  {Fiore}, {Bongiorno}, {Lanzuisi}, {Merloni}, {Schramm}, {Silverman}, \&
  {Civano}}]{Cresci2015a}
{Cresci}, G., {Mainieri}, V., {Brusa}, M., {et~al.} 2015{\natexlab{a}}, \apj,
  799, 82

\bibitem[{{Cresci} {et~al.}(2015{\natexlab{b}}){Cresci}, {Marconi}, {Zibetti},
  {Risaliti}, {Carniani}, {Mannucci}, {Gallazzi}, {Maiolino}, {Balmaverde},
  {Brusa}, {Capetti}, {Cicone}, {Feruglio}, {Bland-Hawthorn}, {Nagao}, {Oliva},
  {Salvato}, {Sani}, {Tozzi}, {Urrutia}, \& {Venturi}}]{Cresci2015b}
{Cresci}, G., {Marconi}, A., {Zibetti}, S., {et~al.} 2015{\natexlab{b}}, \aap,
  582, A63

\bibitem[{{Davies} {et~al.}(2014){Davies}, {Maciejewski}, {Hicks}, {Emsellem},
  {Erwin}, {Burtscher}, {Dumas}, {Lin}, {Malkan}, {M{\"u}ller-S{\'a}nchez},
  {Orban de Xivry}, {Rosario}, {Schnorr-M{\"u}ller}, \& {Tran}}]{Davies2014}
{Davies}, R.~I., {Maciejewski}, W., {Hicks}, E.~K.~S., {et~al.} 2014, \apj,
  792, 101

\bibitem[{{Davies} {et~al.}(2015){Davies}, {Burtscher}, {Rosario},
  {Storchi-Bergmann}, {Contursi}, {Genzel}, {Graci{\'a}-Carpio}, {Hicks},
  {Janssen}, {Koss}, {Lin}, {Lutz}, {Maciejewski}, {M{\"u}ller-S{\'a}nchez},
  {Orban de Xivry}, {Ricci}, {Riffel}, {Riffel}, {Schartmann},
  {Schnorr-M{\"u}ller}, {Sternberg}, {Sturm}, {Tacconi}, \&
  {Veilleux}}]{Davies2015}
{Davies}, R.~I., {Burtscher}, L., {Rosario}, D., {et~al.} 2015, \apj, 806, 127

\bibitem[{{Davies} {et~al.}(2016){Davies}, {Groves}, {Kewley}, {Dopita},
  {Hampton}, {Shastri}, {Scharw{\"a}chter}, {Sutherland}, {Kharb}, {Bhatt},
  {Jin}, {Banfield}, {Zaw}, {James}, {Juneau}, \& {Srivastava}}]{Davies2016}
{Davies}, R.~L., {Groves}, B., {Kewley}, L.~J., {et~al.} 2016, \mnras, 462,
  1616

\bibitem[{{Diniz} {et~al.}(2015){Diniz}, {Riffel}, {Storchi-Bergmann}, \&
  {Winge}}]{Diniz2015}
{Diniz}, M.~R., {Riffel}, R.~A., {Storchi-Bergmann}, T., \& {Winge}, C. 2015,
  \mnras, 453, 1727

\bibitem[{{Durr{\'e}} \& {Mould}(2018)}]{Durre2018}
{Durr{\'e}}, M., \& {Mould}, J. 2018, \apj, 867, 149

\bibitem[{{Durr{\'e}} \& {Mould}(2019)}]{Durre2019}
---. 2019, \apj, 870, 37

\bibitem[{{Emsellem} {et~al.}(2001){Emsellem}, {Greusard}, {Combes}, {Friedli},
  {Leon}, {P{\'e}contal}, \& {Wozniak}}]{Ensellem2001}
{Emsellem}, E., {Greusard}, D., {Combes}, F., {et~al.} 2001, \aap, 368, 52

\bibitem[{{Evans} {et~al.}(2010){Evans}, {Primini}, {Glotfelty}, {Anderson},
  {Bonaventura}, {Chen}, {Davis}, {Doe}, {Evans}, {Fabbiano}, {Galle}, {Gibbs},
  {Grier}, {Hain}, {Hall}, {Harbo}, {He}, {Houck}, {Karovska}, {Kashyap},
  {Lauer}, {McCollough}, {McDowell}, {Miller}, {Mitschang}, {Morgan},
  {Mossman}, {Nichols}, {Nowak}, {Plummer}, {Refsdal}, {Rots}, {Siemiginowska},
  {Sundheim}, {Tibbetts}, {Van Stone}, {Winkelman}, \& {Zografou}}]{Evans2010}
{Evans}, I.~N., {Primini}, F.~A., {Glotfelty}, K.~J., {et~al.} 2010, \apjs,
  189, 37

\bibitem[{{Fabian}(1994)}]{Fabian1994}
{Fabian}, A.~C. 1994, \araa, 32, 277

\bibitem[{{Fabian}(2012)}]{Fabian2012}
---. 2012, \araa, 50, 455

\bibitem[{{Feruglio} {et~al.}(2010){Feruglio}, {Maiolino}, {Piconcelli},
  {Menci}, {Aussel}, {Lamastra}, \& {Fiore}}]{Feruglio2010}
{Feruglio}, C., {Maiolino}, R., {Piconcelli}, E., {et~al.} 2010, \aap, 518,
  L155

\bibitem[{{Fiore} {et~al.}(2017){Fiore}, {Feruglio}, {Shankar}, {Bischetti},
  {Bongiorno}, {Brusa}, {Carniani}, {Cicone}, {Duras}, {Lamastra}, {Mainieri},
  {Marconi}, {Menci}, {Maiolino}, {Piconcelli}, {Vietri}, \&
  {Zappacosta}}]{Fiore2017}
{Fiore}, F., {Feruglio}, C., {Shankar}, F., {et~al.} 2017, \aap, 601, A143

\bibitem[{{Fischer} {et~al.}(2013){Fischer}, {Crenshaw}, {Kraemer}, \&
  {Schmitt}}]{Fischer2013}
{Fischer}, T.~C., {Crenshaw}, D.~M., {Kraemer}, S.~B., \& {Schmitt}, H.~R.
  2013, \apjs, 209, 1

\bibitem[{{Fischer} {et~al.}(2010){Fischer}, {Crenshaw}, {Kraemer}, {Schmitt},
  \& {Trippe}}]{Fischer2010}
{Fischer}, T.~C., {Crenshaw}, D.~M., {Kraemer}, S.~B., {Schmitt}, H.~R., \&
  {Trippe}, M.~L. 2010, \aj, 140, 577

\bibitem[{{Fischer} {et~al.}(2017){Fischer}, {Machuca}, {Diniz}, {Crenshaw},
  {Kraemer}, {Riffel}, {Schmitt}, {Baron}, {Storchi-Bergmann}, {Straughn},
  {Revalski}, \& {Pope}}]{Fischer2017}
{Fischer}, T.~C., {Machuca}, C., {Diniz}, M.~R., {et~al.} 2017, \apj, 834, 30

\bibitem[{{Fischera} \& {Dopita}(2005)}]{Fischera2005}
{Fischera}, J., \& {Dopita}, M. 2005, \apj, 619, 340

\bibitem[{{Flower} \& {Pineau Des For{\^e}ts}(2010)}]{Flower2010}
{Flower}, D.~R., \& {Pineau Des For{\^e}ts}, G. 2010, \mnras, 406, 1745

\bibitem[{{Fluetsch} {et~al.}(2019){Fluetsch}, {Maiolino}, {Carniani},
  {Marconi}, {Cicone}, {Bourne}, {Costa}, {Fabian}, {Ishibashi}, \&
  {Venturi}}]{Fluetsch2019}
{Fluetsch}, A., {Maiolino}, R., {Carniani}, S., {et~al.} 2019, \mnras, 483,
  4586

\bibitem[{{F{\"o}rster Schreiber} {et~al.}(2014){F{\"o}rster Schreiber},
  {Genzel}, {Newman}, {Kurk}, {Lutz}, {Tacconi}, {Wuyts}, {Bandara}, {Burkert},
  {Buschkamp}, {Carollo}, {Cresci}, {Daddi}, {Davies}, {Eisenhauer}, {Hicks},
  {Lang}, {Lilly}, {Mainieri}, {Mancini}, {Naab}, {Peng}, {Renzini}, {Rosario},
  {Shapiro Griffin}, {Shapley}, {Sternberg}, {Tacchella}, {Vergani},
  {Wisnioski}, {Wuyts}, \& {Zamorani}}]{ForsterSchreiber2014}
{F{\"o}rster Schreiber}, N.~M., {Genzel}, R., {Newman}, S.~F., {et~al.} 2014,
  \apj, 787, 38

\bibitem[{{Gadotti} {et~al.}(2019){Gadotti}, {S{\'a}nchez-Bl{\'a}zquez},
  {Falc{\'o}n-Barroso}, {Husemann}, {Seidel}, {P{\'e}rez}, {de
  Lorenzo-C{\'a}ceres}, {Martinez-Valpuesta}, {Fragkoudi}, {Leung}, {van de
  Ven}, {Leaman}, {Coelho}, {Martig}, {Kim}, {Neumann}, \&
  {Querejeta}}]{Gadotti2019}
{Gadotti}, D.~A., {S{\'a}nchez-Bl{\'a}zquez}, P., {Falc{\'o}n-Barroso}, J.,
  {et~al.} 2019, \mnras, 482, 506

\bibitem[{{Gallagher} {et~al.}(2019){Gallagher}, {Maiolino}, {Belfiore},
  {Drory}, {Riffel}, \& {Riffel}}]{Gallagher2019}
{Gallagher}, R., {Maiolino}, R., {Belfiore}, F., {et~al.} 2019, \mnras, 485,
  3409

\bibitem[{{Garc{\'{\i}}a-Burillo} {et~al.}(2014){Garc{\'{\i}}a-Burillo},
  {Combes}, {Usero}, {Aalto}, {Krips}, {Viti}, {Alonso-Herrero}, {Hunt},
  {Schinnerer}, {Baker}, {Boone}, {Casasola}, {Colina}, {Costagliola},
  {Eckart}, {Fuente}, {Henkel}, {Labiano}, {Mart{\'{\i}}n}, {M{\'a}rquez},
  {Muller}, {Planesas}, {Ramos Almeida}, {Spaans}, {Tacconi}, \& {van der
  Werf}}]{Garcia-Burillo2014}
{Garc{\'{\i}}a-Burillo}, S., {Combes}, F., {Usero}, A., {et~al.} 2014, \aap,
  567, A125

\bibitem[{{G{\"u}ver} \& {{\"O}zel}(2009)}]{Guver2009}
{G{\"u}ver}, T., \& {{\"O}zel}, F. 2009, \mnras, 400, 2050

\bibitem[{{Harrison}(2017)}]{Harrison2017}
{Harrison}, C.~M. 2017, Nature Astronomy, 1, 0165

\bibitem[{{Harrison} {et~al.}(2014){Harrison}, {Alexander}, {Mullaney}, \&
  {Swinbank}}]{Harrison2014}
{Harrison}, C.~M., {Alexander}, D.~M., {Mullaney}, J.~R., \& {Swinbank}, A.~M.
  2014, \mnras, 441, 3306

\bibitem[{{Humire} {et~al.}(2018){Humire}, {Nagar}, {Finlez}, {Firpo},
  {Slater}, {Lena}, {Soto-Pinto}, {Mu{\~n}oz-Vergara}, {Riffel}, {Schmitt},
  {Kraemer}, {Schnorr-M{\"u}ller}, {Fischer}, {Robinson}, {Storchi-Bergmann},
  {Crenshaw}, \& {Elvis}}]{Humire2018}
{Humire}, P.~K., {Nagar}, N.~M., {Finlez}, C., {et~al.} 2018, \aap, 614, A94

\bibitem[{{Kang} \& {Woo}(2018)}]{Kang2018}
{Kang}, D., \& {Woo}, J.-H. 2018, \apj, 864, 124

\bibitem[{{Karouzos} {et~al.}(2016{\natexlab{a}}){Karouzos}, {Woo}, \&
  {Bae}}]{Karouzos2016a}
{Karouzos}, M., {Woo}, J.-H., \& {Bae}, H.-J. 2016{\natexlab{a}}, \apj, 819,
  148

\bibitem[{{Karouzos} {et~al.}(2016{\natexlab{b}}){Karouzos}, {Woo}, \&
  {Bae}}]{Karouzos2016b}
---. 2016{\natexlab{b}}, \apj, 833, 171

\bibitem[{{Kauffmann} \& {Heckman}(2009)}]{Kauffmann2009}
{Kauffmann}, G., \& {Heckman}, T.~M. 2009, \mnras, 397, 135

\bibitem[{{Lena} {et~al.}(2015){Lena}, {Robinson}, {Storchi-Bergman},
  {Schnorr-M{\"u}ller}, {Seelig}, {Riffel}, {Nagar}, {Couto}, \&
  {Shadler}}]{Lena2015}
{Lena}, D., {Robinson}, A., {Storchi-Bergman}, T., {et~al.} 2015, \apj, 806, 84

\bibitem[{{Leroy} {et~al.}(2009){Leroy}, {Walter}, {Bigiel}, {Usero}, {Weiss},
  {Brinks}, {de Blok}, {Kennicutt}, {Schuster}, {Kramer}, {Wiesemeyer}, \&
  {Roussel}}]{Leroy2009}
{Leroy}, A.~K., {Walter}, F., {Bigiel}, F., {et~al.} 2009, \aj, 137, 4670

\bibitem[{{Luo} {et~al.}(2016){Luo}, {Hao}, {Blanc}, {Jogee}, {van den Bosch},
  \& {Weinzirl}}]{Luo2016}
{Luo}, R., {Hao}, L., {Blanc}, G.~A., {et~al.} 2016, \apj, 823, 85

\bibitem[{{Maiolino} {et~al.}(2017){Maiolino}, {Russell}, {Fabian}, {Carniani},
  {Gallagher}, {Cazzoli}, {Arribas}, {Belfiore}, {Bellocchi}, {Colina},
  {Cresci}, {Ishibashi}, {Marconi}, {Mannucci}, {Oliva}, \&
  {Sturm}}]{Maiolino2017}
{Maiolino}, R., {Russell}, H.~R., {Fabian}, A.~C., {et~al.} 2017, \nat, 544,
  202

\bibitem[{{Markwardt}(2009)}]{Markwardt2009}
{Markwardt}, C.~B. 2009, in Astronomical Society of the Pacific Conference
  Series, Vol. 411, Astronomical Data Analysis Software and Systems XVIII, ed.
  D.~A. {Bohlender}, D.~{Durand}, \& P.~{Dowler}, 251

\bibitem[{{Meijerink} {et~al.}(2013){Meijerink}, {Kristensen}, {Wei{\ss}}, {van
  der Werf}, {Walter}, {Spaans}, {Loenen}, {Fischer}, {Israel}, {Isaak},
  {Papadopoulos}, {Aalto}, {Armus}, {Charmandaris}, {Dasyra}, {Diaz-Santos},
  {Evans}, {Gao}, {Gonz{\'a}lez-Alfonso}, {G{\"u}sten}, {Henkel}, {Kramer},
  {Lord}, {Mart{\'{\i}}n-Pintado}, {Naylor}, {Sanders}, {Smith}, {Spinoglio},
  {Stacey}, {Veilleux}, \& {Wiedner}}]{Meijerink2013}
{Meijerink}, R., {Kristensen}, L.~E., {Wei{\ss}}, A., {et~al.} 2013, \apjl,
  762, L16

\bibitem[{{Mullaney} {et~al.}(2013){Mullaney}, {Alexander}, {Fine}, {Goulding},
  {Harrison}, \& {Hickox}}]{Mullaney2013}
{Mullaney}, J.~R., {Alexander}, D.~M., {Fine}, S., {et~al.} 2013, \mnras, 433,
  622

\bibitem[{{M{\"u}ller-S{\'a}nchez} {et~al.}(2011){M{\"u}ller-S{\'a}nchez},
  {Prieto}, {Hicks}, {Vives-Arias}, {Davies}, {Malkan}, {Tacconi}, \&
  {Genzel}}]{Muller-Sanchez2011}
{M{\"u}ller-S{\'a}nchez}, F., {Prieto}, M.~A., {Hicks}, E.~K.~S., {et~al.}
  2011, \apj, 739, 69

\bibitem[{{Murphy} {et~al.}(2011){Murphy}, {Condon}, {Schinnerer}, {Kennicutt},
  {Calzetti}, {Armus}, {Helou}, {Turner}, {Aniano}, {Beir{\~a}o}, {Bolatto},
  {Brandl}, {Croxall}, {Dale}, {Donovan Meyer}, {Draine}, {Engelbracht},
  {Hunt}, {Hao}, {Koda}, {Roussel}, {Skibba}, \& {Smith}}]{Murphy2011}
{Murphy}, E.~J., {Condon}, J.~J., {Schinnerer}, E., {et~al.} 2011, \apj, 737,
  67

\bibitem[{{Rakshit} \& {Woo}(2018)}]{Rakshit2018}
{Rakshit}, S., \& {Woo}, J.-H. 2018, \apj, 865, 5

\bibitem[{{Revalski} {et~al.}(2018){Revalski}, {Crenshaw}, {Kraemer},
  {Fischer}, {Schmitt}, \& {Machuca}}]{Revalski2018}
{Revalski}, M., {Crenshaw}, D.~M., {Kraemer}, S.~B., {et~al.} 2018, \apj, 856,
  46

\bibitem[{{Riffel} {et~al.}(2009){Riffel}, {Storchi-Bergmann}, {Dors}, \&
  {Winge}}]{Riffel2009}
{Riffel}, R.~A., {Storchi-Bergmann}, T., {Dors}, O.~L., \& {Winge}, C. 2009,
  \mnras, 393, 783

\bibitem[{{Riffel} {et~al.}(2014){Riffel}, {Storchi-Bergmann}, \&
  {Riffel}}]{Riffel2014}
{Riffel}, R.~A., {Storchi-Bergmann}, T., \& {Riffel}, R. 2014, \apjl, 780, L24

\bibitem[{{Riffel} {et~al.}(2013){Riffel}, {Storchi-Bergmann}, \&
  {Winge}}]{Riffel2013}
{Riffel}, R.~A., {Storchi-Bergmann}, T., \& {Winge}, C. 2013, \mnras, 430, 2249

\bibitem[{{Riffel} {et~al.}(2008){Riffel}, {Storchi-Bergmann}, {Winge},
  {McGregor}, {Beck}, \& {Schmitt}}]{Riffel2008}
{Riffel}, R.~A., {Storchi-Bergmann}, T., {Winge}, C., {et~al.} 2008, \mnras,
  385, 1129

\bibitem[{{Schommer} {et~al.}(1988){Schommer}, {Caldwell}, {Wilson}, {Baldwin},
  {Phillips}, {Williams}, \& {Turtle}}]{Schommer1988}
{Schommer}, R.~A., {Caldwell}, N., {Wilson}, A.~S., {et~al.} 1988, \apj, 324,
  154

\bibitem[{{Sch{\"o}nell} {et~al.}(2019){Sch{\"o}nell}, {Storchi-Bergmann},
  {Riffel}, {Riffel}, {Bianchin}, {Dahmer-Hahn}, {Diniz}, \&
  {Dametto}}]{Schonell2019}
{Sch{\"o}nell}, A.~J., {Storchi-Bergmann}, T., {Riffel}, R.~A., {et~al.} 2019,
  \mnras, 485, 2054

\bibitem[{{Silk}(2013)}]{Silk2013}
{Silk}, J. 2013, \apj, 772, 112

\bibitem[{{Silk} \& {Rees}(1998)}]{Silk1998}
{Silk}, J., \& {Rees}, M.~J. 1998, \aap, 331, L1

\bibitem[{{Slater} {et~al.}(2019){Slater}, {Nagar}, {Schnorr-M{\"u}ller},
  {Storchi-Bergmann}, {Finlez}, {Lena}, {Ramakrishnan}, {Mundell}, {Riffel},
  {Peterson}, {Robinson}, \& {Orellana}}]{Slater2019}
{Slater}, R., {Nagar}, N.~M., {Schnorr-M{\"u}ller}, A., {et~al.} 2019, \aap,
  621, A83

\bibitem[{{Son} {et~al.}(2009){Son}, {Hyung}, {Ferruit}, {P{\'e}contal}, \&
  {Lee}}]{Son2009a}
{Son}, D.-H., {Hyung}, S., {Ferruit}, P., {P{\'e}contal}, E., \& {Lee}, W.-B.
  2009, \mnras, 395, 692

\bibitem[{{Storchi-Bergmann} {et~al.}(2010){Storchi-Bergmann}, {Lopes},
  {McGregor}, {Riffel}, {Beck}, \& {Martini}}]{Storchi-Bergmann2010}
{Storchi-Bergmann}, T., {Lopes}, R.~D.~S., {McGregor}, P.~J., {et~al.} 2010,
  \mnras, 402, 819

\bibitem[{{van der Werf} {et~al.}(2010){van der Werf}, {Isaak}, {Meijerink},
  {Spaans}, {Rykala}, {Fulton}, {Loenen}, {Walter}, {Wei{\ss}}, {Armus},
  {Fischer}, {Israel}, {Harris}, {Veilleux}, {Henkel}, {Savini}, {Lord},
  {Smith}, {Gonz{\'a}lez-Alfonso}, {Naylor}, {Aalto}, {Charmandaris}, {Dasyra},
  {Evans}, {Gao}, {Greve}, {G{\"u}sten}, {Kramer}, {Mart{\'{\i}}n-Pintado},
  {Mazzarella}, {Papadopoulos}, {Sanders}, {Spinoglio}, {Stacey}, {Vlahakis},
  {Wiedner}, \& {Xilouris}}]{vanderWerf2010}
{van der Werf}, P.~P., {Isaak}, K.~G., {Meijerink}, R., {et~al.} 2010, \aap,
  518, L42

\bibitem[{{Vazdekis} {et~al.}(2016){Vazdekis}, {Koleva}, {Ricciardelli},
  {R{\"o}ck}, \& {Falc{\'o}n-Barroso}}]{Vazdekis2016}
{Vazdekis}, A., {Koleva}, M., {Ricciardelli}, E., {R{\"o}ck}, B., \&
  {Falc{\'o}n-Barroso}, J. 2016, \mnras, 463, 3409

\bibitem[{{Veilleux} {et~al.}(2013){Veilleux}, {Mel{\'e}ndez}, {Sturm},
  {Gracia-Carpio}, {Fischer}, {Gonz{\'a}lez-Alfonso}, {Contursi}, {Lutz},
  {Poglitsch}, {Davies}, {Genzel}, {Tacconi}, {de Jong}, {Sternberg}, {Netzer},
  {Hailey-Dunsheath}, {Verma}, {Rupke}, {Maiolino}, {Teng}, \&
  {Polisensky}}]{Veilleux2013}
{Veilleux}, S., {Mel{\'e}ndez}, M., {Sturm}, E., {et~al.} 2013, \apj, 776, 27

\bibitem[{{Vogt} {et~al.}(2013){Vogt}, {Dopita}, \& {Kewley}}]{Vogt2013}
{Vogt}, F.~P.~A., {Dopita}, M.~A., \& {Kewley}, L.~J. 2013, \apj, 768, 151

\bibitem[{{Wilson} {et~al.}(1993){Wilson}, {Braatz}, {Heckman}, {Krolik}, \&
  {Miley}}]{Wilson1993}
{Wilson}, A.~S., {Braatz}, J.~A., {Heckman}, T.~M., {Krolik}, J.~H., \&
  {Miley}, G.~K. 1993, \apjl, 419, L61

\bibitem[{{Woo} {et~al.}(2016){Woo}, {Bae}, {Son}, \& {Karouzos}}]{Woo2016}
{Woo}, J.-H., {Bae}, H.-J., {Son}, D., \& {Karouzos}, M. 2016, \apj, 817, 108

\bibitem[{{Woo} {et~al.}(2017){Woo}, {Son}, \& {Bae}}]{Woo2017}
{Woo}, J.-H., {Son}, D., \& {Bae}, H.-J. 2017, \apj, 839, 120

\bibitem[{{Zinn} {et~al.}(2013){Zinn}, {Middelberg}, {Norris}, \&
  {Dettmar}}]{Zinn2013}
{Zinn}, P.-C., {Middelberg}, E., {Norris}, R.~P., \& {Dettmar}, R.-J. 2013,
  \apj, 774, 66

\bibitem[{{Zschaechner} {et~al.}(2016){Zschaechner}, {Walter}, {Bolatto},
  {Farina}, {Kruijssen}, {Leroy}, {Meier}, {Ott}, \&
  {Veilleux}}]{Zschaechner2016}
{Zschaechner}, L.~K., {Walter}, F., {Bolatto}, A., {et~al.} 2016, \apj, 832,
  142

\bibitem[{{Zubovas} \& {Bourne}(2017)}]{Zubovas2017}
{Zubovas}, K., \& {Bourne}, M.~A. 2017, \mnras, 468, 4956

\bibitem[{{Zubovas} {et~al.}(2013){Zubovas}, {Nayakshin}, {King}, \&
  {Wilkinson}}]{Zubovas2013}
{Zubovas}, K., {Nayakshin}, S., {King}, A., \& {Wilkinson}, M. 2013, \mnras,
  433, 3079

\end{thebibliography}

\end{document}